\documentclass[20pt,twocolumn]{IEEEtran}
\usepackage{amsmath}
\usepackage{amsthm,amssymb,amsmath,bm}
\usepackage{subfigure}
\usepackage{amsfonts}
\usepackage{epsfig}
\usepackage{algcompatible}
\usepackage{amssymb}
\usepackage{color}
\usepackage{transparent}
\usepackage{graphicx}
\usepackage{amsmath}
\graphicspath{{image/}}
\usepackage{epstopdf}
\usepackage[table,xcdraw]{xcolor}
\usepackage{cite}
\usepackage[utf8x]{inputenc}
\usepackage{color,soul}
\usepackage{graphicx}
\usepackage{epstopdf}
\usepackage{graphics}
\usepackage{subfigure}
\usepackage{multirow}
\usepackage{rotating}
\usepackage{graphicx}
\usepackage{tabularx}
\usepackage{array}
\usepackage{color,soul}
\usepackage{bm}
\usepackage{graphicx,dblfloatfix}
\usepackage{blindtext}

\DeclareMathOperator*{\argmax}{arg\,max} 
\usepackage{subfigure}
\usepackage{moreverb}
\usepackage{epsfig}
\usepackage{amsmath,amssymb,amsthm,mathrsfs,amsfonts,dsfont}
\usepackage{epstopdf}
\usepackage{adjustbox,lipsum}
\usepackage{amsfonts}
\usepackage{epsfig}
\usepackage{amssymb}
\usepackage{amsmath}
\usepackage{amsthm}
\usepackage{subfigure}
\usepackage{multirow}
\usepackage{rotating}
\usepackage{graphicx}
\usepackage{tabularx}
\usepackage{array}
\usepackage{anyfontsize}
\usepackage{color,soul}
\usepackage{graphicx,dblfloatfix}
\usepackage{epstopdf}
\usepackage{blindtext}
\usepackage{amsmath}
\usepackage{amsthm,amssymb,amsmath,bm}
\usepackage{subfigure}
\usepackage{amsfonts}
\usepackage{epsfig}
\usepackage{amssymb}
\usepackage{amsfonts}
\usepackage{amsmath}
\usepackage{cite}
\usepackage{hyperref}
\usepackage{graphicx}
\usepackage{fancyhdr}
\usepackage{subfigure}
\usepackage[subfigure]{tocloft}
\usepackage[font={small}]{caption}
\usepackage{subfigure}
\usepackage{tabularx}
\usepackage{tcolorbox}
\usepackage{cite}
\usepackage[linesnumbered,ruled,vlined]{algorithm2e}
\SetKwInput{KwInput}{Input}
\SetKwInput{KwOutput}{Output}
\usepackage{amsthm,amssymb,amsmath,bm}
\usepackage{graphicx}
\usepackage{fancyhdr}
\usepackage{subfigure}
\usepackage[subfigure]{tocloft}
\usepackage[font={small}]{caption}
\usepackage{subfigure}
\usepackage{tabularx}
\usepackage{cite}
\allowdisplaybreaks
\usepackage{blindtext}

\IEEEdisplaynontitleabstractindextext
\IEEEpeerreviewmaketitle
\begin{document}
\title{Online Service Provisioning  in
		NFV-enabled Networks
		Using Deep Reinforcement Learning}	
\author{\IEEEauthorblockN{Ali Nouruzi, Abolfazl Zakeri, \IEEEmembership{Student  Member, IEEE}, Mohamad Reza Javan, \IEEEmembership{Senior Member, IEEE},  Nader Mokari, \IEEEmembership{Senior Member, IEEE}, Rasheed Hussain, \IEEEmembership{Senior Member, IEEE}, Ahsan Syed Kazmi, \IEEEmembership{Student Member, IEEE}
	\thanks{	
		A. Nouruzi, 
		A. Zakeri and N. Mokari are with the Department of Electrical and Computer Engineering,~Tarbiat Modares University,~Tehran,~Iran, (e-mail: nader.mokari@modares.ac.ir). M.~R.~Javan is with the Department of Electrical Engineering, Shahrood University of Technology, Iran, (e-mail: javan@shahroodut.ac.ir).~R. Hussain and A. S. Kazmi are with the Institute of Information Security and Cyber-Physical Systems, Innopolis University, Russia, (e-mail: r.hussain@innopolis.ru).
}}}
\maketitle
\begin{abstract}
In this paper, we study a Deep Reinforcement Learning (DRL) based framework for  an online end-user service provisioning in a Network Function Virtualization (NFV)-enabled network.  \textcolor{black}{We formulate an optimization problem aiming  to minimize the cost of network resource utilization}. The main challenge is provisioning  the online service requests by fulfilling their Quality of Service (QoS)   
under limited resource availability. Moreover, fulfilling the stochastic service requests in a large  network is  another challenge that is evaluated in this paper.  
To solve the formulated optimization problem in an efficient and intelligent manner, \textcolor{black}{we propose a Deep Q-Network for Adaptive Resource allocation (DQN-AR) in NFV-enable network    for function placement and dynamic routing which considers the available network resources as DQN states.} Moreover, the service's characteristics, including the service life time and number of the arrival requests, are modeled by  the  Uniform and Exponential distribution, respectively. In addition, we evaluate the computational complexity of the proposed method. 
\textcolor{black}{Numerical results carried out for different  ranges of parameters  reveal the effectiveness of our framework}. In specific,  the obtained results show that the average  number of admitted requests of the network increases by 7 up to 14$\%$ and the network utilization cost decreases by 5 and 20$\%$.
	\end{abstract}
	\section{INTRODUCTION}
	\subsection{State of The Art and Motivation
	 }
In recent years, new applications have emerged rapidly with diverse  Quality of Service	(QoS) requirements \cite{series2015imt}. To meet their requirements in an efficient manner with a common physical infrastructure, exploiting advanced technologies is indispensable where these technologies are expected to have pivotal impacts on network performance in terms of enhancing QoS and resource efficiency which result in cost reduction. One such technology is Network Function Virtualization (NFV)  providing an array of benefits such as great flexibility, resource efficiency, and cost reduction  \cite{cziva2017container}.
However, in such an NFV-enabled network, providing an efficient resource allocation algorithm is a challenging task. In addition, handle online service requests and also service arrival and  departure and  its effect on the network resources are the other challenges that we have in this paper.  To tackle these challenges and design adaptive and intelligent networks, recently, Deep Reinforcement Learning (DRL) based methods have been used to solve various resource allocation problems\cite{8403657, 8382166}.
 \\\indent Besides, devising online and adaptive/on-demand service provisioning algorithms  under dynamic network resource variations is  another challenging task in NFV-enabled  networks. Online and on-demand services by considering lifetime, i.e,    
while previously provisioned  services are running, new  service requests can arrive. 
Recently, some researchers have made great efforts   to address the mentioned challenges, but, to the best of our knowledge, a few researchers consider the lifetime and online service requests, and the effect of this, on the Resource Allocation (RA) problem \cite{herrera2016resource, hamann2019path, 10.1145/3326285.3329056}.  The prevailing works  inspire us to
seek a ``smart`` and ``online'' service provisioning  method with considering service ``lifetime'' in a NFV-enabled network. The term ``service'' indicates a type of end user request with specific QoS and Service Function Chain (SFC) characteristics, and  ``provisioning'' means that such request's requirements are fulfilled, hence,  the service request is admitted. In brief, this work focuses on a main question which is: \textit{how a service provider offers heterogeneous services with a probabilistic lifetime on the common physical resources in a smart and efficient  manner?}

	\subsection{{Research Outputs and Contributions}}
	\textcolor{black}{Different from previous works \cite{zhang2019deep,10.1145/3326285.3329056,gholipoor2020e2e}, this work provides a DRL-based online service provisioning algorithm in an NFV-enabled network	in which the considered service requests are online with a probabilistic lifetime. In addition, by deploying the proposed online	service provisioning method, new requests can be served while	previously accepted services are running.}\\
The main results and contributions of this work are
listed as follows:
\\$\bullet$~We propose a new service assurance model  leveraging NFV  to perform their Network Functions (NFs) and guarantee  QoS  in terms of latency and bandwidth. To this end, we formulate an optimization problem with constraints on the QoS and the limitation of network resources.
\\$\bullet$~
We deploy DRL method
 to fulfill the different requests at each time under dynamic network resources  in a long-term run.
  To improve the convergence speed of the considered case with large number  of states and large action space, we deploy a Deep Q-Network (DQN) algorithm.  
\\$\bullet$~We propose an online service provisioning method in which each user has a service timeline and new users can request service at each time slot while some services are running from the previous time slots. We also consider the dynamics of resource consumption and release in the network due to admitting new services and terminating previous services. Moreover, request arrivals are modeled by Uniform distribution, and service duration time (service time) is modeled by the exponential distribution. 
\\$\bullet$~To apply the DQN algorithm for RA (i.e., solving the optimization problem), we develop the available calculation algorithm that updates the state space (e.g., due to resource releasing or failure occurring) at the beginning of each time slot to find an appropriate action. 
\\$\bullet$~To evaluate the performance of the proposed method, we consider different baselines.
The obtained results unveil  that the proposed method has considerable performance. The main baselines are greedy and online-Tabu search algorithms which  are well known methods in online algorithm consideration. 

\subsection{\textcolor{black}{Related Work}}
Recently some works define a service with a specific SFC which includes a set of Virtual Network Functions (VNFs)  and these VNFs need to be executed in a tolerable delay \cite{alleg2017delay}, \cite{ren2020efficient}. These VNFs run on a specific virtual machines which are created on  top of the physical network by leveraging NFV.  Hereupon,  service provisioning means that the requested SFC with QoS requirement for each request is done successfully by performing the SFC  and RA in the NFV environment \cite{kamgang2020slice}.
The basic principles of NFV Resource Allocation (NFV-RA) is studied in \cite{herrera2016resource} comprehensively. Also, online scheduling with minimizing the total execution time of VNFs is studied in \cite{mijumbi2015design}. Furthermore,
the authors in \cite{hamann2019path} propose NFV-RA for traffic routing by deploying game theory. They focus on routing and embedding of VNFs and do not consider the scheduling problem. Similarly, placement of VNF instances for different services with link allocation and fixed delay for links is studied in \cite{nguyen2019practical}. Delay-aware cost minimization for random arrival service requests by deploying stochastic dual
gradient method is studied in \cite{chen2018multi}.

At the same time,
 DRL-based methods to solve various RA problems have attracted much attention \cite{mao2018deep, li2018deep, ayoubi2018machine}. 
 In \cite{zorello2018improving}, a ML algorithm for extracting feature of data traffic in NFV-cloud network for predicting  computation and demands of resources is deployed. 
In \cite{ding2019deep},  DRL based mechanism with Markov Decision Process (MDP) is proposed for reducing congestion probability and also choosing  transmission path for routing and traffic engineering. Network congestion probability reduce to $50\%$ with compare to Open Short First Path (OSFP) routing method. Because  routing and function placement problems  are related to each other, \cite{pei2018virtual} proposes a function placement and chaining schemes, jointly with  Binary Integer Programming (BIP) for minimizing End to End (E2E) delay, and then use Restricted Boltzmann Machine (RBM) output to  determine the next hope node in the network. The authors in \cite{zhou2019multi} proposes  multi-task deep learning for routing and dynamic SFC with considering network status for predicting  the routing path. Lastly, in \cite{subramanya2019machine}, the  authors use Integer Linear Programming (ILP) and multi layer perceptron  to minimize E2E delay and placement of VNFs.  
 In \cite{wu2019resource}, the authors  propose multi-objective programming and assume that access points  work as a player in a game theory based problem that minimizes OpEx and average response time.
 \textcolor{black}{In \cite{fu2019dynamic},the authors study providing IoT services in an NFV-enabled network by deploying DRL. Aiming to minimize the processing and transition delay, the proposed DRL method reduces the total delay to around $200$ms that  has decreased up to 3 times compared to other baselines. In \cite{pham2017virtual}, the anthers propose a matching-based scheduling method  that reduces the scheduling time in a NFV-enabled network up to $50\%$  compared to the Round-Robin scheduling method. Aiming to provide a cost-efficient dynamic resource management in a NFV-enabled network, the authors in \cite{li2020finedge} propose a practical method that reduces CPU utilization up to $10\%$  compared to the traditional approaches. In addition, the authors in \cite{ning2020deep, pei2019optimal, qu2020dynamic} study the performance of the DRL-based methods for   RA in the context of a NFV-enabled network where th obtained results show a significant improvement in the results obtained results  compared to the traditional optimization methods. Motivated by significant effectiveness of DRL-based algorithm for RA in NFV-enabled networks, we propose a DRL-based algorithm for service provision in an NFV-enabled network. In addition, concerning the ability of DRL to support online algorithms, the proposed DQN algorithm is adopted an online RA algorithm that  different from previous works \cite{jia2018online,huang2019maximizing,xu2018throughput}, we assume that the services arrive based on the real stochastic model. Moreover, the required resources are allocated to the services while the subsequent services arrive.}
We summarize  related works and compare them with our work in Table. \ref{Main_Com}.  
	\begin{table*}[h!]
	\centering
	\caption{\textcolor{black}{Related Works Summary}}
	\scalebox{.7}{
		\begin{tabular}{|
				>{\columncolor[HTML]{96FFFB}}c |c|c|c|c|}
			\hline
			\multicolumn{1}{|l|}{\cellcolor[HTML]{34CDF9}Ref.} & \cellcolor[HTML]{34CDF9}Scenarios                                          & \cellcolor[HTML]{34CDF9}Strategy                                                                                      & \cellcolor[HTML]{34CDF9}Main Contribution                                                                                     & \cellcolor[HTML]{34CDF9}Differences with this work                                                                                                                                                                                                                                                      \\ \hline
			\cite{ding2019deep}                                      & \begin{tabular}[c]{@{}c@{}}Routing and traffic\\  engineering\end{tabular} & \begin{tabular}[c]{@{}c@{}}Using DQN\\ for Routing algorithm\end{tabular}                                            & \begin{tabular}[c]{@{}c@{}}Proposing an online routing \\ for Routing algorithm\end{tabular}                                  & \begin{tabular}[c]{@{}c@{}}Considering nodes with several \\ VMs and services specification\end{tabular}                                                         \\\hline
			\cite{fu2019dynamic}                                     & \begin{tabular}[c]{@{}c@{}}NFV \\ (SFC \& Routing)\end{tabular}             & \begin{tabular}[c]{@{}c@{}}SFC embedding for\\ NFV-enabled IoT\\ and routing by shortest path algorithm\end{tabular} & Dynamic SFC embedding                                                                                                         & \begin{tabular}[c]{@{}c@{}}Dynamic routing algorithm  by considering links state\end{tabular}\\ \hline
			\cite{pham2017virtual}     & \begin{tabular}[c]{@{}c@{}}NFV \\ (SFC \& Scheduling)\end{tabular}          & \begin{tabular}[c]{@{}c@{}}Matching-Based \\ VNF Scheduling\end{tabular}  & \begin{tabular}[c]{@{}c@{}}RA to VNF over time\\ with a matching scheme\end{tabular} & \begin{tabular}[c]{@{}c@{}}Online service assurance with \\ considering VM and links  and using DQN\end{tabular}                                                                                     \\ \hline
			\cite{zhou2019multi}                                     & \begin{tabular}[c]{@{}c@{}}NFV\\  (SFC \& Routing)\end{tabular} & Multi task Deep learning  & \begin{tabular}[c]{@{}c@{}}Learning  traffic routing by SFC\\ information\end{tabular}    & \begin{tabular}[c]{@{}c@{}}Objective, function placement\\  with considering  VM states and dynamic  routing algorithm  \\ by considering links state\end{tabular}                                \\ \hline
			\cite{pei2018virtual}                                    & \begin{tabular}[c]{@{}c@{}}NFV \\ (SFC \& Routing)\end{tabular}        
			& \begin{tabular}[c]{@{}c@{}}Applying Deep learning (RBM) \\ to solve a BIP\end{tabular}                                & \begin{tabular}[c]{@{}c@{}}Minimizing E2E delay \\ with considering SFC path
			\end{tabular}                                     & \begin{tabular}[c]{@{}c@{}}Objective, VM state function placement \\ consideration and link's state dynamic routing
			\end{tabular}             
			\\ \hline
\cite{li2020finedge}                                    &        NFV (SFC)
			&           Deploying testbed                    &  \begin{tabular}[c]{@{}c@{}} Real-time flow monitoring and\\ dynamic resource management  \end{tabular}         &   \begin{tabular}[c]{@{}c@{}}    Real service consideration and\\ evaluation of service life time    \end{tabular} 
			\\ \hline
\cite{ning2020deep}                                    &  \begin{tabular}[c]{@{}c@{}} NFV  \\(SFC \& Routing) \end{tabular}    
	&        \begin{tabular}[c]{@{}c@{}}      Deploying  DRL to solve\\ MILP for optimize resource utilization \end{tabular}              &      Near optimal results  is obtained         &          Evaluation of effect of network topology and geo-distributed DC
	\\ \hline
\cite{10.1145/3326285.3329056}	& NFV (SFC)&Deploying  DRL to function placement &  Deploying DRL for SFC  & \begin{tabular}[c]{@{}c@{}}Dynamic routing and \\ considering real service characteristics \end{tabular}
		\\ \hline
		\cite{pei2019optimal}	&NFV (SFC \& Routing)&Deploying DRL for solving BIP&Dynamic SFC embedding& Objective and node by node dynamic routing
		\\ \hline
		\cite{qu2020dynamic}	&NFV (SDN \& Routing)&\begin{tabular}[c]{@{}c@{}}Deploying DRL to solve mixed\\ integer quadratic constrained
			(MIQCP) programming\end{tabular}&Real time traffic model and NF migration& Objective and service life time consideration and evaluation of network topology
		\\ \hline
\end{tabular}
	}
	\label{Main_Com}
\end{table*}
\subsection{{Paper Organization}}
This paper is arranged as follows:  Section \ref{systemmodelandproblemformulation} displays the proposed system model and problem formulation. Section \ref{solution} presents the solution methods of the formulated problem. Computational complexity of the proposed algorithm and baselines is evaluated in Section \ref{section_complexity}. Simulation results are provided in Section \ref{Simulation_Results}. At the end, concluding remarks are stated in Section \ref{Conclusion}.
\\\indent
\textbf{Symbol Notations:} 
 We use $\lfloor .\rfloor$  for representing floor function, that takes  input and  gives the greatest integer less than or equal to the input. $|.|$ denotes the absolute value or size of input argument and $a_i$ shows the $i$-th element of vector $\bold{a}$ and $a_{i,j}$ shows the $i,j$ element of matrix $\bold{A}$.	Also to define a set and its elements, we use $\mathcal{B}$ and $b_{n}$ respectively where $b_n$ is the $n$-th elements of $\mathcal{B}$. We use ${\Bbb{R}}_{+}$  and $\Bbb{N}$ to show the set of positive real numbers and natural numbers, receptively.  In addition, for representing  modulo operation for  the remainder of the  division of $a$ by $n$, we use $a\equiv_n$. 
	\section{{PROPOSED SYSTEM MODEL AND PROBLEM FORMULATION}}\label{systemmodelandproblemformulation}
\begin{figure}
	\centering
	\includegraphics[width=0.9\linewidth]{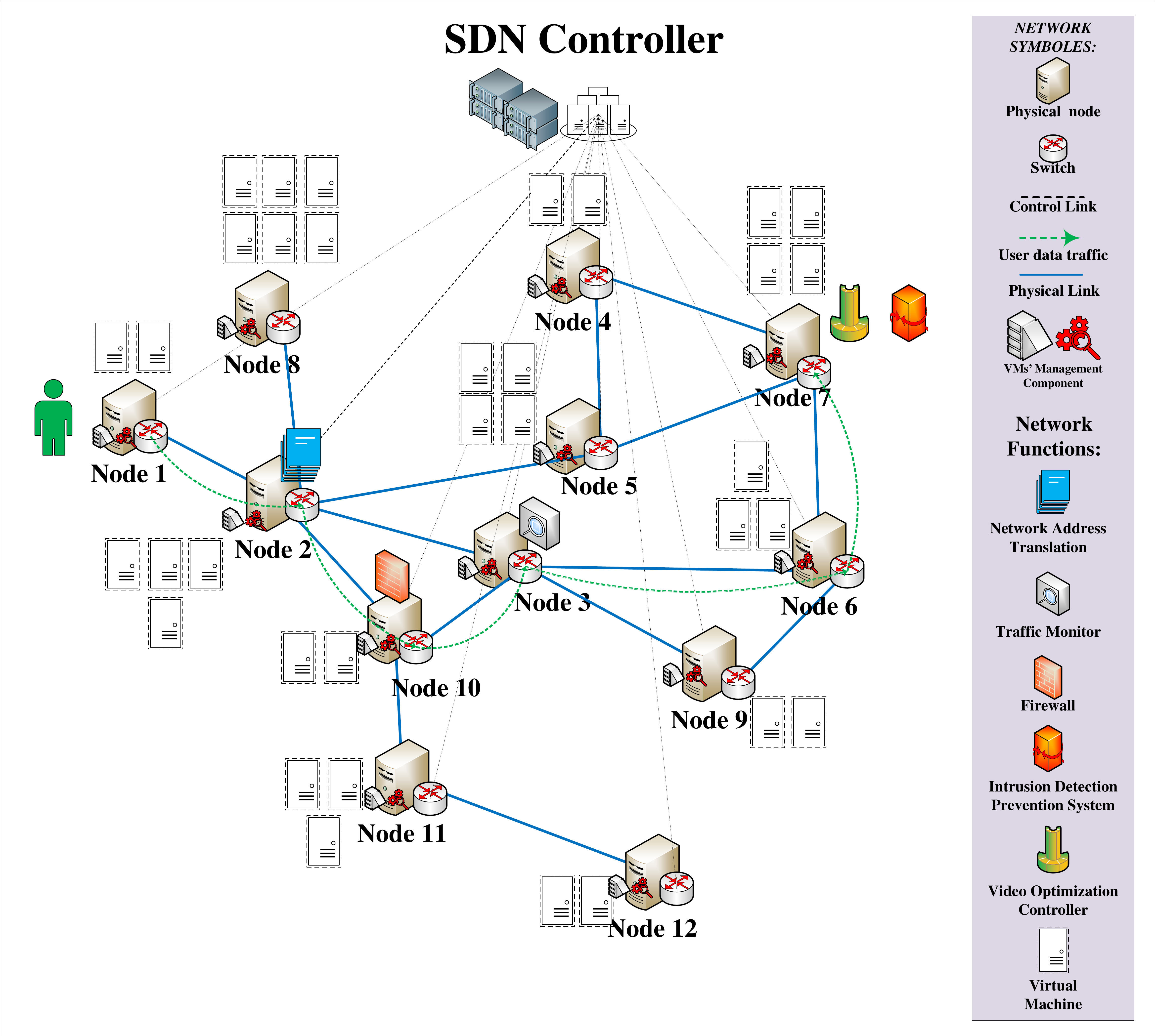}
		\caption{High level representation of the considered system model. }
\label{fig:mec-nfv-task-service}
\end{figure}

\textcolor{black}{The proposed system model that has two  parts: 1) user's request with service characteristics and requirements}
 and 2) NFV-enabled infrastructure, and an optimization problem for allocating the resources of the infrastructure to the services. We assume a central controller for providing cooperation and coordination between the network component, and  a software-based network control. The high-level representation of the proposed system model is  depicted in Fig. \ref{fig:mec-nfv-task-service}. More details about this figure are provided in the following subsection.
	\subsection{Service Specification and Requirements}
\begin{figure}
	\centering
	\includegraphics[width=0.7\linewidth]{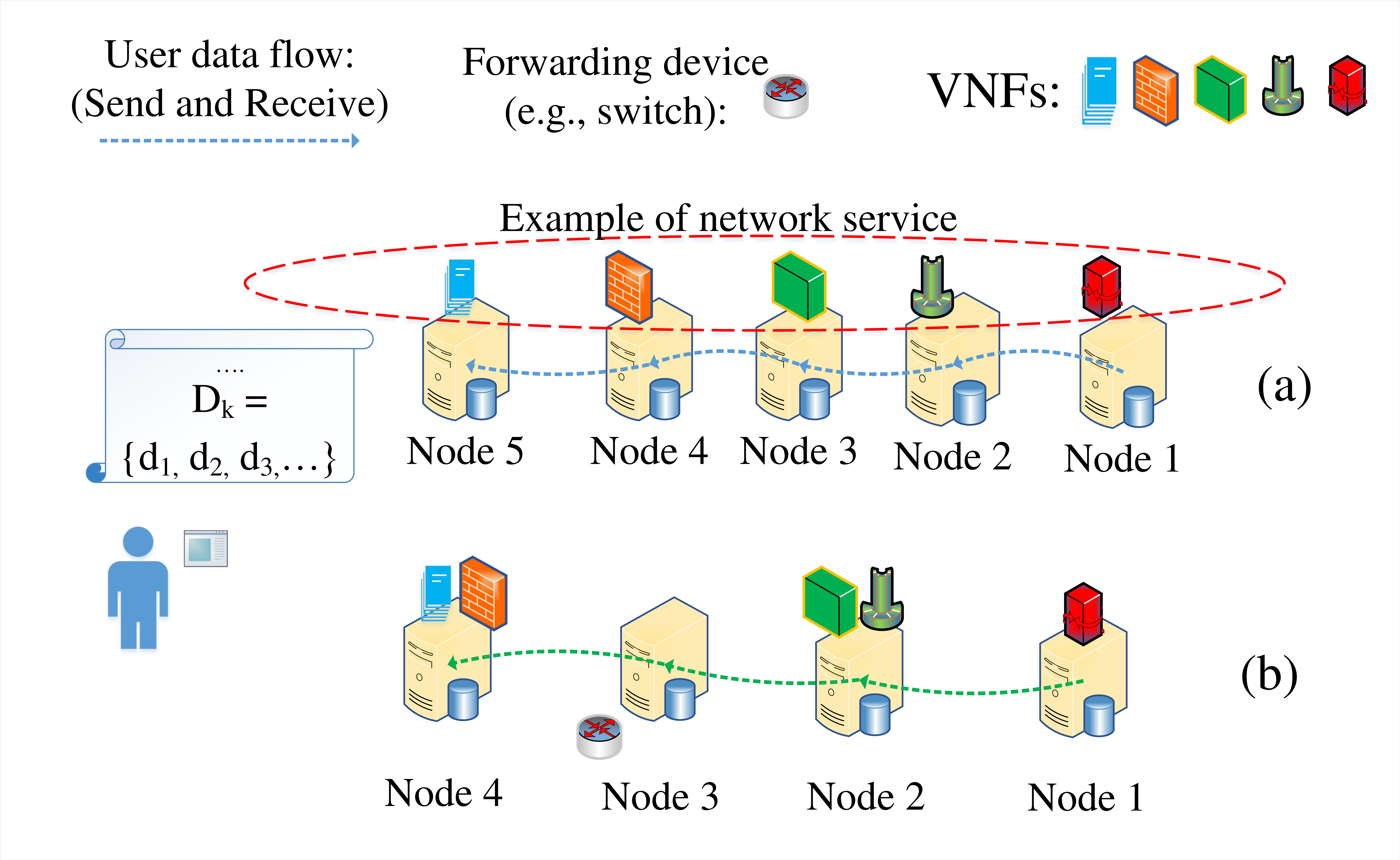}
			\caption{An example of SFC with different scenarios for function placement. We assume that the functions can placed on the  successive nodes (a) or  non successive nodes (b). }
\label{fig:SFC}
\end{figure}

Based on the 3rd Generation Partnership Project (3GPP) standardization perspective \cite{thirdgeneration2017system}, each communication service needs some NFs that run on the flow/packets of the services. European Telecommunications Standards Institute (ETSI) defines a set of NFs with specific chaining and descriptors as a Network Service (NS) \cite{etsi2014gs} in the NFV environments.  
According to these, we consider a set of $K$ services which is denoted by $\mathcal{K}=\{1,\dots, K\}$ and a set of all NFs as $\mathcal{F}=\{1,\dots,F\}$. Each service $k$ has some NFs with specific ordering as an SFC that is shown in Fig. \ref{fig:SFC}. We assume that  $\mathcal{F}_k \subset\mathcal{F}$ is the set of specific functions of service $k$ like Firewall (FW), Network Address Translator (NAT), Intrusion Detection Prevention System (IDPS), and Video Optimization Controller (VOC). We assume that each service $k$ is specified by following:
 \begin{align}
 R_k=\Big(n_{i,k},n_{e,k}, B_k,\tau_k,\mathcal{D}_k\Big), \forall k\in \mathcal{K},
 \end{align} 
 where $n_{i,k}$ and $n_{e,k}$ are the ingress and egress nodes of service $k$ \cite{pei2018virtual,hong2018resource} and $\left\lbrace n_{i,k},n_{e,k}\right\rbrace \in\mathcal{N}$. \textcolor{black}{It is worth mentioning that each of the services has a specific sequence of functions. For example, in the VoIP service, FW  runs after NAT} \cite{savi2019impact}. In addition, $B_k$ is the  data rate in bits per second. Moreover,	$\tau_k$ is the tolerable time which is dependent on the type of services of the top layer with respect to their latency requirements.\footnote{Note that $\tau_{k}$ is not the E2E latency and is the SFC latency. Hence, it is\ the latency of the core network in the view of the cellular network.} Also, we define $d^{k}_f$ to determine the corresponding  processing requirement for virtualized NF (VNF)  $f$ in CPU cycle per bits of flow/packet in service $k$ \cite{savi2019impact}. Accordingly, for each service, we have a set of corresponding processing requirements as bellow:
 \footnote{Obviously, the layer two and layer three NFs have different characteristics and requirements as layer-2/3 processing in \cite{liu2018microboxes}.}  
 \begin{align}
 	\mathcal{D}_{k}=\{d_{{f}}^k\}, ~~~~\forall f\in\mathcal{F}_{k}, \forall k \in \mathcal{K}.
 	\label{D_k}
 \end{align}
	\begin{table}[h!]
		\renewcommand{\arraystretch}{0.8}
		\centering
		\caption{Main notations and parameters}
		\label{System_parameters}
			\scalebox{.6}{
		\begin{tabular}{|c|l| }	
			\hline
			\rowcolor[HTML]{38FFF8} 
			\textbf{Notation}& \textbf{Definition}\\\hline
			$\mathcal{G(N,L)}$&Network graph\\ \hline
			$\mathcal{N}/n$&Set/index of nodes\\\hline
			$\mathcal{L}$&Set of links\\\hline
			$\mathcal{U}/u$&Set/index of users\\\hline
			$\mathcal{F}/f$&Set/index of  NFs\\\hline
			$\mathcal{F}_k$&Set NFs of service $k$\\\hline
			$\mathcal{V}/v$&Set/index of  VMs\\\hline
			$\mathcal{P}/p$	& Set/index of the physical paths\\\hline
			$\mathcal{E}/e$	& Set/index of the virtual paths\\\hline
			$\boldmath{L}$ & Connectivity matrix of graph\\\hline
			$x_{n,n'} $ &Capacity of the link between nodes $n$ and $n'$ in bit per second\\\hline
			$w_{m,v}$ &Weight/unit costs of VM $v$ on node $m$  \\ \hline
			$\hat{w}_{n,n'}$ &Weight/unit costs of the link between nodes $n$ and $n'$
            \\\hline
			$n_{i,k}, n_{e,k}$&The ingress and egress nodes of service $k$  \\\hline
			$i^{p_{m,m'}}_{n,n'}$&Link indicator that shows that link between nodes\\& $n$ and $n'$ is placed on the physical path $p_{m,m'}$  \\\hline
			$B_k$&Data rate for service $k$ in bits per second \\\hline
			$\tilde{B}_k$&Packet size in bits\\\hline
			$\mathcal{D}_k$&Set of the corresponding processing \\& requirement in CPU cycle per bit \\& for the functions of service $k$\\\hline
			$d_{f}^{k}$&Corresponding processing requirement in CPU cycle per bit \\& for function $f$ of service $k$\\\hline
			$\tau_k$ & Tolerable latency of service  $k$\\\hline
			$\xi_{v,m}^{f,u} \in \left\lbrace0,1 \right\rbrace $& Selection indicator of VM $v$ for NF $f$ on node $m$ for user $u$\\\hline
			$\gamma$&Decay factor of reinforcement learning\\\hline
			$\alpha$&Learning rate for DQN \\\hline
			$\rho^{k,e^{v,v'}_{m,m'}}_{{p}_{m,m'}}\in \left\lbrace0,1 \right\rbrace $& Path selection variable that mapping  the virtual\\& path $e$ between virtual machine\\& $v$ and $v'$ for service $k$  to physical path $p_{n,n'}$\\& between nodes $n$ and $n'$\\\hline
			$\Psi_{v,m}$&Processing resource at VM $v$ \\& on nodes $m$ in CPU cycle per second\\\hline 
			$\delta_{u}^k$& Service request indicator where set to $1$  \\&for user $u$ that requests service $k$  \\\hline
			${z}_{v,m}^{t}$& Available processing resource of VM $v$ that is \\&raised on node $m$ at time slot $t$\\\hline
			${y}_{{n,n'}}^{t}$&Available capacity resource of link between \\& nodes $n$ and $n'$ at time slot $t$\\\hline
		\end{tabular}
	}
	\end{table}
Also, to determine the order of the successive functions in  a certain SFC, we define the order  of   functions by $f^{i}$ and $f^{i+1}$, where $f^{i}$ is  $i$-th function of the SFC and  $f^{i+1}$ is run after function $f^{i}$. To increase the readability of this paper, the main parameters and variables are summarized in Table \ref{System_parameters}.
Moreover, we consider a set $\mathcal{U}$ of users with different service requests. We assume that each user $u$  requests only one service. We define a binary indicator  $\delta_{u}^{k}$, where if user $u$ requests service $k$, it is $1$ and otherwise  $0$. 
	\subsection{{Infrastructure Model}}
	In order to model and formulate the NFV-enabled network, we  consider graph  $\mathcal{G}=(\mathcal{N},\mathcal{L})$, where $\mathcal {N}$ represents the set of  nodes where $\left| \mathcal{N}\right|=N $ and $\mathcal{L}$ is the set of links between nodes. 
 We further assume that each node $m$  hosts several  VMs  that is denoted by $\mathcal{V}_{m}=\{1_m,\dots,V_m\}$ and created by a hypervisor, hence the set of total VMs  in the network is denoted by $\mathcal{V_\text{Total}}=\cup_{m=1}^{N}\mathcal{V}_m$.
 In addition, we denote the  maximum number of the VMs on each nodes by $V_\text{max}$.
	\\ \indent
Each VM $v$ on node $m$ has a specific processing resource  that is denoted by $\Psi_{v,m}$ in CPU cycle per second. Hence, matrix $\boldsymbol{\Psi}=[\Psi_{v,m}]\in \Bbb{R}_{+}^{V\times N}$ 
indicates the amount of processing resources and also determine the VMs of each node.  It is possible that each VM processes a set of NFs for different users based on the allowable capacity \cite{gholipoor2020e2e}.
 Moreover,  we consider connectivity matrix as $\bold{L}=\left[  l_{n,n'}\right] $, that is defined as
	\begin{equation}
	\begin{split}
	&l_{n,n'}= \begin{cases}
	1, & \text{a link between nodes $n$ and $n'$ exists},\\
	0, & \text{otherwise}.
	\end{cases}
	\end{split} 
	\end{equation}
Also, the link between nodes $n$ and $n'$ has a limited bandwidth that is represented by matrix \textcolor{black}{$\bold{X}= [x_{n,n'}]\in{\Bbb{R}}_{+}^{N\times N}$, where $x_{n,n'}$ is the capacity of link  between nodes ${n}$ and ${n'}$ in bits per second.}	Note that as the considered network is connected, there is at least one path between two nodes. Let $p_{m,m'}$ denotes the
$p$-th path between nodes $m$ and $m'$. Therefore,  we have a set $\mathcal{P}_{m,m'}=\left\{1_{m,m'},\dots,p_{m,m'},\dots,P_{m,m'}\right\}$ of all possible physical paths between nodes $m$ and $m'$ such that each path  contains a set of links. To determine which of  the physical links are in a path, we define a link-to-path binary indicator as follows:
\begin{align}
i^{p_{m,m'}}_{n,n'}=	\begin{cases}
1, & \text{the link between nodes $n$ and $n'$} \\&\text{  is in the path $p_{m,m'},$}
\\
0, & \text{otherwise}.
\end{cases}
\end{align}	
Moreover,  we consider the set of virtual paths  between virtual  machine $v$  and $v'$  on nodes $m$ and $m'$ as  $\mathcal{E}^{v,v'}_{m,m'}=\left\lbrace 1^{v,v'}_{m,m'},\dots,e^{v,v'}_{m,m'}, \dots, E^{v,v'}_{m,m'} \right\rbrace $  where  $e^{v,v'}_{m,m'}$ is the  e-th path of this set \cite{tajiki2018joint}, \cite{ebrahimi2020joint}, and \cite{miotto2019adaptive}. \footnote{ In addition, we assume that in each of physical nodes, there are unlimited bandwidth links between the VMs. Moreover, we assume that there is at least a physical path for each virtual path.}
\subsection{Optimization Variables}
We define a binary decision variable $\xi_{v,m}^{f^{i},k}$ to determine that $i$-th function of service $k$ is running on VM $v$ that is raised on node $m$ as follows:
\begin{align}
\xi_{v,m}^{f^{i},k}=\begin{cases}
1,& \text{  NF $f^{i}$ of service $k$ is running on VM $v$ in} \\&\text{  node $m$,} \\0,& \text{otherwise}.
\end{cases} 
\end{align} 
Moreover, to send data traffic of service $k$, we define a binary decision variable $\rho^{k,e^{v,v'}_{m,m'}}_{p_{m,m'}}$ where it maps the virtual path $e^{v,v'}_{m,m'}$ to the physical path $p_{m,m'}$ as follows:
\begin{align}
\rho^{k,e^{v,v'}_{m,m'}}_{p_{m,m'}}=
\begin{cases}
1,& \text{the physical path $p_{m,m'}$ is selected to tarnsmit  }\\& \text{ of service the data traffic of service $k$ from} \\&\text{ from virtual machine $v$ to $v'$},\\
0,& \text{otherwise},
\end{cases}
\end{align}
where for each virtual path just one physical path is selected. Based on this,  we define the following constraint:
\begin{align}
\label{sum1}
\sum_{p_{m,m'}\in \mathcal{P}_{m,m'}}\rho^{k,e^{v,v'}_{m,m'}}_{p_{m,m'}}=1, \forall k \in \mathcal{K}.
\end{align}
We note that the virtual path $e^{v,v'}_{m,m'}$ is between two successive functions of SFC of service $k$, $(f^{i},f^{i+1})$, with respect to the ordering of SFC. For example, in a certain service, the functions like web browsing, NAT function are always run before FW.\\
Moreover, the path between  $n_{i,k}$ and the VM that the first function  of SFC is placed is determined by $\hat{e}^{v}_{n_{i,k},n}$ and also for the path between the VM that  the last function placed on it and $n_{e,k}$ is determined by $\hat{e}^{v}_{n,n_{e,k}}$.
\subsection{Delay Model}
This work considers three types of delays as: 1) processing delay, 2) propagation delay  and 3) transmission delay.
\subsubsection{Processing delay}
The processing delay of NF $f$ on node $m$ for service $k$ in VM $v$  denoted by $\tau_{v,m}^{f,k}$  in seconds  is given by
	\begin{align}
	\tau_{v,m}^{f,k}
	=\frac{d_{{f}}^{k}\tilde{B}_{k}}{\Psi_{v,m}},\forall k\in\mathcal{K},  v\in\mathcal{V}_m, m \in \mathcal{N},
	\end{align}
	where  $\tilde{B}_k$ is the packet size in bits. In this paper, we assume the packet size is equal to the number of bits transmitted in one second. For example, by considering a service with required $100$ Kbps data rate, the packet size is $100$ Kbits \cite{gholipoor2020e2e}. Also, the total of processing delay of service  $k$ can be calculated by 
	\begin{align}
	D_{\text{Proc}}^{k}=\sum^{F_k}_{i=1}\sum_{m\in\mathcal{N}}\sum_{v\in\mathcal{V}_{m}}\xi_{v,m}^{f^{i},k}\tau_{v,m}^{f^{i},k}, \forall k \in \mathcal{K}, f^{i}\in\mathcal{F}_k,\in\mathcal{K}.
	\end{align}
\subsubsection{{ Propagation Delay}}
To formulate the propagation delay in the considered system, we define $\kappa_{n,n'}$
as  the amount of propagation delay for the data traffic that traverses on link between nodes $n$ and $n'$  depends on the length of  this link and the speed of light. Therefore, the total propagation delay for service $k$ is obtained by:
\begin{align}
\label{prob}
&D^{k}_{\text{Prop}}=
\sum_{\substack{{n,n',m}\in\mathcal{N}\\ p_{n_{i,k},m'}\in\mathcal{P}_{n_{i,k},m'}\\v\in\mathcal{V}_m}}
 \kappa_{n,n}i^{p_{n_{i,k},m}}_{n,n'}\rho^{\hat{e}^{v}_{n_{i,k},m}}_{p_{n_{i,k},m}}\xi^{f^{1},k}_{v,m}+\\
\nonumber
&\sum^{F_k-1}_{i=1}\sum_{\substack {{n,n',m',m''}\in\mathcal{N}\\p_{m',m''}\in\mathcal{P}_{m',m''}\\ v',v''\in\mathcal{V}_m}}
 \kappa_{n,n'}i^{p_{m',m''}}_{n,n'}\rho^{e^{v',v''}_{m',m''}}_{p_{m',m''}}\xi^{f^{i},k}_{v',m'}\xi^{f^{i+1},k}_{v'',m''}+\\
 \nonumber
&
\sum_{\substack{n,n',m'''\in\mathcal{N}\\ p_{m''',n_{e,k}}\in\mathcal{P}_{m''',n_{e,k}}\\{v'''\in\mathcal{V}_m}}}\kappa_{{n,n'}} i^{p_{m''',n_{e,k}}}_{{n,n'}}\rho^{\hat{e}^{v'''}_{m''',n_{e,k}}}_{p_{m''',n_{e,k}}}\xi^{f^{F_k},k}_{v''',m'''},\\& \nonumber \forall k\in\mathcal{K}.
\end{align}
In the first term of \eqref{prob}, we calculate the propagation delay between $n_{i,k}$ and the first VM that the first  function is placed. In addition, the second term calculate the propagation delay of the link between the next functions. Finally the last term calculates the propagation delay on the link between the last VM that and $n_{e,k}$.
\subsubsection{Transmission Delay}
	The total transmission delay of service $k$ is calculated by:
	\begin{figure}[h!]
		\centering
		\includegraphics[width=1\linewidth]{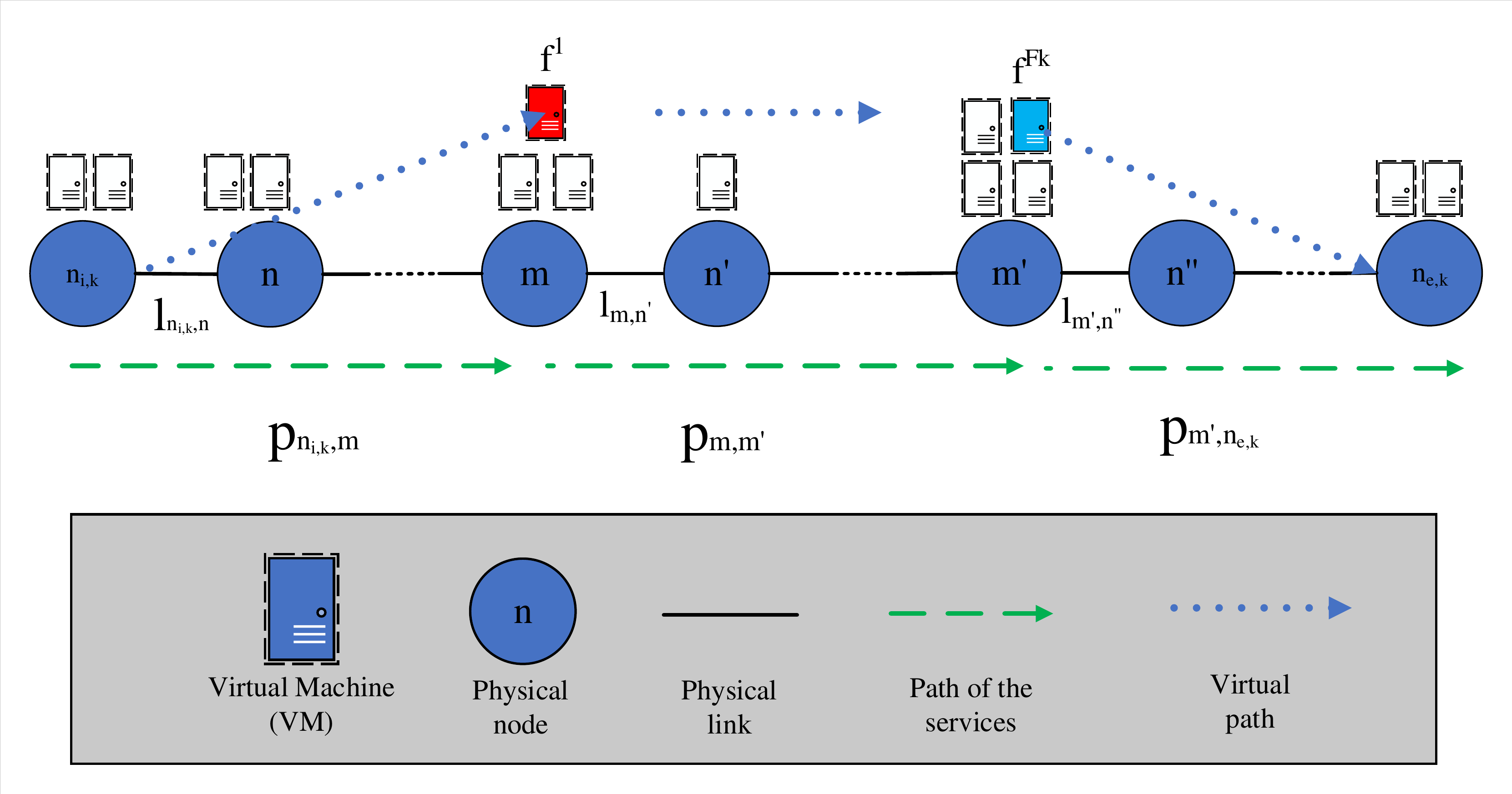}
		\caption{The illustration of equations \eqref{prob} and \eqref{trans} for calculation of propagation and transmission delay. These equations are included three terms that calculate the propagation and transmission delay between the ingress nodes and the first VM' node and the last VM's node and egress node of services.  }
		\label{fig:rout1}
	\end{figure}
\begin{align}
\label{trans}
&D_{\text{Tran}}^{k}=
\sum_{\substack{{n,n',m}\in\mathcal{N}\\
p_{n_{i,k},m}\in\mathcal{P}_{n_{i,k},m}\\{v\in\mathcal{V}_m}}}i^{p_{n_{i,k},m}}_{n,n'}\rho^{k,\hat{e}^{v}_{n_{i,k},m}}_{p_{n_{i,k},m}}\xi^{f^{1},k}_{v,m}\frac{\tilde{B}_{k}}{x_{n,n'}}+
\nonumber	\\ &\sum^{F_k-1}_{i=1}
\sum_{\substack{{n,n',m',m''}\in\mathcal{N}\\ p_{m',m''}\in\mathcal{P}_{m',m''}\\{v',v''\in\mathcal{V}_m}}}i^{p_{m',m''}}_{{n,n'}}\rho^{k,e^{v',v''}_{m',m''}}_{p_{m',m''}}\xi^{f^{i},k}_{v',m'}\xi^{f^{i+1},k}_{v'',m''}\frac{\tilde{B}_{k}}{x_{n,n'}}+
\nonumber \\ &\sum_{\substack{n,n',m'''\in\mathcal{N}
\\p_{m''',n_{e,k}}\in\mathcal{P}_{m''',n_{e,k}}\\v'''\in\mathcal{V}_m}}i^{p_{m''',n_{e,k}}}_{{n,n'}}\rho^{k,\hat{e}^{v'''}_{m''',n_{e,k}}}_{p_{m''',n_{e,k}}}\xi^{f^{F_k},k}_{v''',m'''}\frac{\tilde{B}_{k}}{x_{n,n'}}
, \\ \nonumber
&\forall k\in\mathcal{K}, \forall f^{i}\in\mathcal{F}_k.
	\end{align}
  To have better understanding and realization of \eqref{trans} and \eqref{prob}, the details of these equations are depicted in Fig. \ref{fig:rout1}.
Based on the formulated latency, the total delay for each packet of service $k$ is obtained as:
\begin{align}
D_{\text{Total}}^k=D_{\text{Proc}}^{k}+D_{\text{Prop}}^{k}+D^{k}_{\text{Tran}}, \forall k \in \mathcal{K}.
\end{align}
\subsection{Objective Function}	
\textcolor{black}{We define a weighted  cost function that includes the  cost  of processing  and bandwidth  resources at the level of VMs and links that is given by:} 

\begin{align}
\label{main_problem}
&\phi(\boldsymbol{\rho}, \boldsymbol{\xi}) =\sum_{k\in\mathcal{K}}\sum_{\substack{{u\in\mathcal{U}}\\{m \in\mathcal{N}}\\{v\in\mathcal{V}_m}\\{f\in\mathcal{F}_k}}} w_{m,v}d_{{f}}^{k}{B}_{k}\delta_{u}^{k}\xi_{v,m}^{f,k} +\\
\nonumber &\sum_{k\in\mathcal{K}}\sum_{\substack{{u\in\mathcal{U}}\\{n,n',m,m'\in\mathcal{N}} \\p_{m,m'}\in\mathcal{P}_{m,m'}\\v,v'\in \mathcal{V}_m}} \hat{w}_{{n,n'}}\delta^{k}_{u} i^{p_{m,m'}}_{n,n'}\rho^{k,e^{v,v'}_{m,m'}}_{p_{m,m'}}\xi^{f,k}_{v,m}\xi^{f',k}_{v',m'}B_k,
\end{align}
\textcolor{black}{
where $w_{m,v} > 0$  denotes the  unit cost of VM $v$ on node $m$ that converts the utilized resources to the cost. 
By considering the service bandwidth and the processing requirement for each of the functions that are placed in the VMs, the total processing cost is calculated by the first term.
Subsequently, $\hat{w}_{n,n'} > 0$ denotes the  unit cost of the link between nodes $n$ and $n'$. By considering the links that are included in the selected paths and the bandwidth of the requested services, the total bandwidth utilization cost is calculated by the second term. \\
The values of parameters $w_{m,n}$ and $\hat{w}_{n,n'}$  depend on the type of nodes, and links, for example, the edge or core nodes has different  (cost) weights.}
Based on the definitions, our main aim is to solve the following optimization problem:
	\begin{subequations}\label{main_prob}
	\begin{align}
		&	\min_{\boldsymbol{\rho}, \boldsymbol{\xi}}\;\phi 
		\\	\textbf{s.t.}~~&  \label{capa_link} 
	 \sum_{m,m'\in\mathcal{N}} i^{p_{m,m'}}_{{n,n'}}\rho^{k,e^{v,v'}_{m,m'}}_{p_{m,m'}}B_k\le x_{n,n'}, \forall k \in\mathcal{K},\\& \nonumber \forall v,v'\in\mathcal{V}_m,\forall n,n'\in\mathcal{N},
		\\&
	\sum_{m\in\mathcal{N}}	\xi^{f^{i},k}_{v,m}d^f_kB_k\le \Psi_{v,m},  f^i\in\mathcal{F}_k,\forall k\in\mathcal{K}, \forall v\in\mathcal{V}_{m},
		\label{capa_cpu}
		\\&  		
		\label{one function}
		\sum_{m\in\mathcal{N}}	\sum_{v\in\mathcal{V}_{m}}\xi_{v,m}^{f^i,k}= 1,~\forall k\in\mathcal{K}, f\in\mathcal{F}_k,
		\\&
		D_{\text{Total}}^k\le \tau_k, ~~~~~~~\forall k\in\mathcal{K}, \label{delay}
		\\&
		\xi_{v,m}^{f^{i},k}\in\{0,1\},~~~~~  v\in \mathcal{V}_{m},  f \in \mathcal{F}_{k},  m\in \mathcal{N}, \forall k \in \mathcal{K},
		\\&
		\rho^{k,e^{v,v'}_{m,m'}}_{{p}_{m,m'}}\in \left\lbrace 0,1\right\rbrace, ~~~ \forall
		k\in\mathcal{K}, p_{m,m'}\in\mathcal{P}_{m,m'}, 
	\end{align}
\end{subequations}
where $\boldsymbol{\rho}=[\rho^{k,e^{v,v'}_{m,m'}}_{{p}_{m,m'}}]$ and $\boldsymbol{\xi}=[\xi_{v,m}^{f,k}]$.
Constraint \eqref{capa_link} ensures that the total resources allocated to service $k$ in all links in path $p_{m,m'}$ are less that the link capacity. Constraint \eqref{capa_cpu} ensures that the total resources  allocated to all users are less than the processing capacity of VM $v$ on node $m$. Constrain \eqref{one function} indicates  that each NF is assigned to one VM. By \eqref{delay}, we consider that the total delay is less than the predefined  tolerable latency of the services.
\section{{PROPOSED SOLUTION }}\label{solution}
\textcolor{black}{Problem \eqref{main_prob} is a  integer linear problem witch is complicated to solve efficiently.  Therefore, we adopt an RL-based algorithm to solve it. Adopting a RL-based solution for solving problem \eqref{main_prob} is a challenge that has significant effect on the obtained  results. In this section, first, we  evaluate the basic principles of  RL algorithms, and second, we describe how to adopt these principles  to solve the proposed problem.}
\subsection{Proposed DQN Adaptive Resource (DQN-AR) Allocation Algorithm}  
We propose a RL-based RA algorithm  with considering the basic concepts of RL. The basics of RL are agent, state, action, reward, and an environment. The agent in each iteration, with considering the state of the environment, selects an action that causes that the state changes into the next state. Subsequently, to evaluate the performance of each action, the agent gets a reward from the environment. The set of states, actions, rewards and next state is collocated in each step of RL based algorithm to the agent, so that based on these experiments, the agent can select better actions in the same states. 
\textcolor{black}{Based on the mentioned assumptions, the main equation for the $Q$-learning algorithm is defined as follows \cite{sutton2018reinforcement}: 
	\begin{align}
	&Q(s^t,a^t)\leftarrow Q(s^t,a^t)+ \\ \nonumber
	&\alpha \left[  r^t+\gamma \argmax_{a'}\left(Q(s^{t+1},a')-Q(s^t,a^t) \right) \right] ,
	\end{align}
	where $s^t$, $a^t$, and $r^t$ denote the state, action, and the obtained reward in the $t$-th step, respectively. In addition, the learning rate and discount factor are denoted by $\alpha$ and $\gamma$, respectively. Because deploying $Q$-learning for the huge state-action space is not possible \cite{li2018deep}, \cite{qu2020dynamic}, a DNN is deployed for estimating the $Q$-function values.}
\\
Based on the  mentioned above, we consider the network components as the basics  of components RL.
\\\textcolor{black}{\textbf{Descriptions of DQN:}
We adopt Algorithm  \ref{DQN}  where the DQN algorithm chooses a random action with probability $\epsilon$. The parameter $\epsilon$ is set to $1$ in the first iteration and  has a final value, $\epsilon=0.1$ whereas the decay coefficient of epsilon is set to $0.9$.  To make sure that the algorithm does not get the local optimum, in each time slot with probability $0 \le \epsilon < 1$, we choose a random action \cite{tokic2011value}. In fact, $\epsilon$  parameters determine the ratio between exploration and exploitation in the search algorithm \cite{sutton2018reinforcement}. In addition, we store the current sate, action, new state, and reward in $\mathcal{\hat{D}}$ memory with a certain size. To update the parameters of  DQN, we sample the set $\hat{\mathcal{B}}$  of the  transactions with the number $|\hat{\mathcal{B}}|$. We set the memory size $|\hat{\mathcal{D}}|=2000$ for storing transactions and 
the size of mini-batch $\hat{\mathcal{B}}$ is set to  $8$  transactions \cite{fu2019dynamic}.  
The learning rate $\alpha$ and the discount factor $\gamma$ is set to  $0.001$ and $0.95$, receptively \cite{10.1145/3326285.3329056,tokic2011value}. 
The reason for using a discount factor $\gamma$ is that it prevents the total reward from going to infinity \cite{van2007reinforcement}.} \\
$\bullet$~\textbf{Agent:} We consider the SDN controller as the agent that by  considering the network's states,  chooses the actions form  action spaces. For each selected action, the agent gets a reward and the network' state changes to  the next state over the time. To have a smart and adaptive algorithm, the agent needs to have knowledge about the network state and condition in each time slot $t$. For this reason, available resources or capacity of nodes and links at each time slot $t$ is necessary \cite{pei2019optimal}. To this end, we propose a available calculation algorithm that more details follow in Algorithm
\ref{Remain_Al}.\\
$\bullet$~\textbf{Network States:} We denote the state space at each time slot $t$  by $\boldsymbol{S}^{t}$ as  network resources
that includes the available resources in terms of processing resources of VMs and links' bandwidth as follows:	
\begin{align}
&\boldsymbol{S}^{t}=(\bold{Z}^{t},\bold{Y}^{t}),\\ \nonumber &\bold{Z}^{t}=[z_{v,m}^{t}]\in \Bbb{R}_{+}^{V\times N},	&\bold{Y}^{t}=[y_{{n,n'}}^{t}]\in\Bbb{R}_{+}^{N\times N},
\end{align}
where $z_{v,m}^{t}$ and $y_{n,n'}^{t}$ are the available processing resource of VM $v$ on physical node $m$ and bandwidth of link between nodes ${n}$ and ${n'}$ in time slot $t$, respectively, and obtained by Algorithm \ref{Remain_Al}. First, we divide the amount of each resource to $I$ levels. To represent the resources state, we normalized the gap between  beginning time and  time slot $t$ as bellow \cite{pei2019optimal}:
   	\begin{align}
     {s}^{t}_{{n,n'}}=\lfloor I\frac{y_{{n,n'}}^{0}-y_{{n,n'}}^{t}}{y_{{n,n'}}^{0}}\rfloor,  ~~~~~ s_{v,n}^{t}=\lfloor I \frac{z_{v,n}^{0}-z_{v,n}^{t}}{z_{v,n}^{0}} \rfloor.
   \end{align}
\textcolor{black}{In order to apply resources' state to input of the DQN, the values of each network component (links and VMs) are normalized. 
 In addition, we set $I$ to 1000 \cite{10.1145/3326285.3329056}}. 
In addition, in RA algorithm, the agent considers the service specification, the previous selected node and VM in the path from ingress to egress nodes, and order of the function in SFC as state. Moreover, we assume that in each  time slot $t$, the agent has some steps to choose action and perform the RA algorithm. We denote the state and action at time slot $t$ and step $j$, for service $k$, by $s^{t,j}$ and  $a^{t,j}_{k}$, respectively. To ensure a limited solving time in each time slot, we assume a upper bound for the steps that is denoted by $J$ and it is set to $100$ in each time slot.  \\
$\bullet~$\textbf{Calculating Available Resources}:\label{Remained}
As mentioned before, we need to have an algorithm that returns the available resources at each time slot. Based on service duration time, the service of users is terminated and their resources are released. Also, to calculate the available resources, it is outlined in Algorithm \ref{Remain_Al}.
\begin{algorithm}[h!]
		\tiny
		\DontPrintSemicolon
		\renewcommand{\arraystretch}{0.9}
		\caption{Resource Allocation and Calculating of Available Resources}
		\label{Remain_Al}
		\KwInput{The network graph and capacity of the links and VMs; $\mathcal{G}$, $\boldsymbol{\Psi}$, and $\bold{X}$}
		\For { each time slot $t$}{
	\For { each services $k$}
	{
		\For { each users  $u$}
		{Save the arrival time ${t}_u$ for user $u$ \\
			\If{the request accepted (By the Actions)}
			{ \If{at the begining time}{$z_{v,m}^{t_u}=\Psi_{v,m}-d_{f}^{k} B_{k}\xi_{v,m}^{f,k}$\\
					$y^{t_u}_{{n,n'}}=x_{n,n'}-B_{k}i^{p_{m,m'}}_{n,n'}\rho^{k,e^{v,v'}_{m,m'}}_{p_{m,m'}}$\\
					\Else{$z_{v,m}^{t_u}=z_{v,m}^{t_{u}-\hat{t}}-d_{f}^{k} B_{k}\xi_{v,m}^{f,k}$\\
						$y^{t_u}_{{n,n'}}=y^{t_u-\hat{t}}_{{n,n'}}+B_{k}i^{p_{m,m'}}_{n,n'}\rho^{k,e^{v,v'}_{m,m'}}_{p_{m,m'}}$\\}
\If{ the user departure}
{\text{Release the user $u$ resource's}
						\\	$z_{v,m}^{t}=z_{v,m}^{t_u}+d_{f}^{k} B_{k}\xi_{v,m}^{f,k}$\\
						$y^{t}_{n,n'}=y^{t_{u}}_{{n,n'}}+B_{k}i^{p_{m,m'}}_{n,n'}\rho^{k,e^{v,v'}_{m,m'}}_{p_{m,m'}}$\\
						${t}_u\leftarrow0$\\}
					\textit{\textbf{Update State}} :$\bold{Z}^t$, $\bold{Y}^t$ according to the utilization of links and nodes based on the bandwidth and processing resources\\}}}}
		\KwOutput{$\bold{Z}^t,\, \bold{Y}^t$} }
\end{algorithm}

\begin{figure}
	\centering
	\includegraphics[width=1\linewidth]{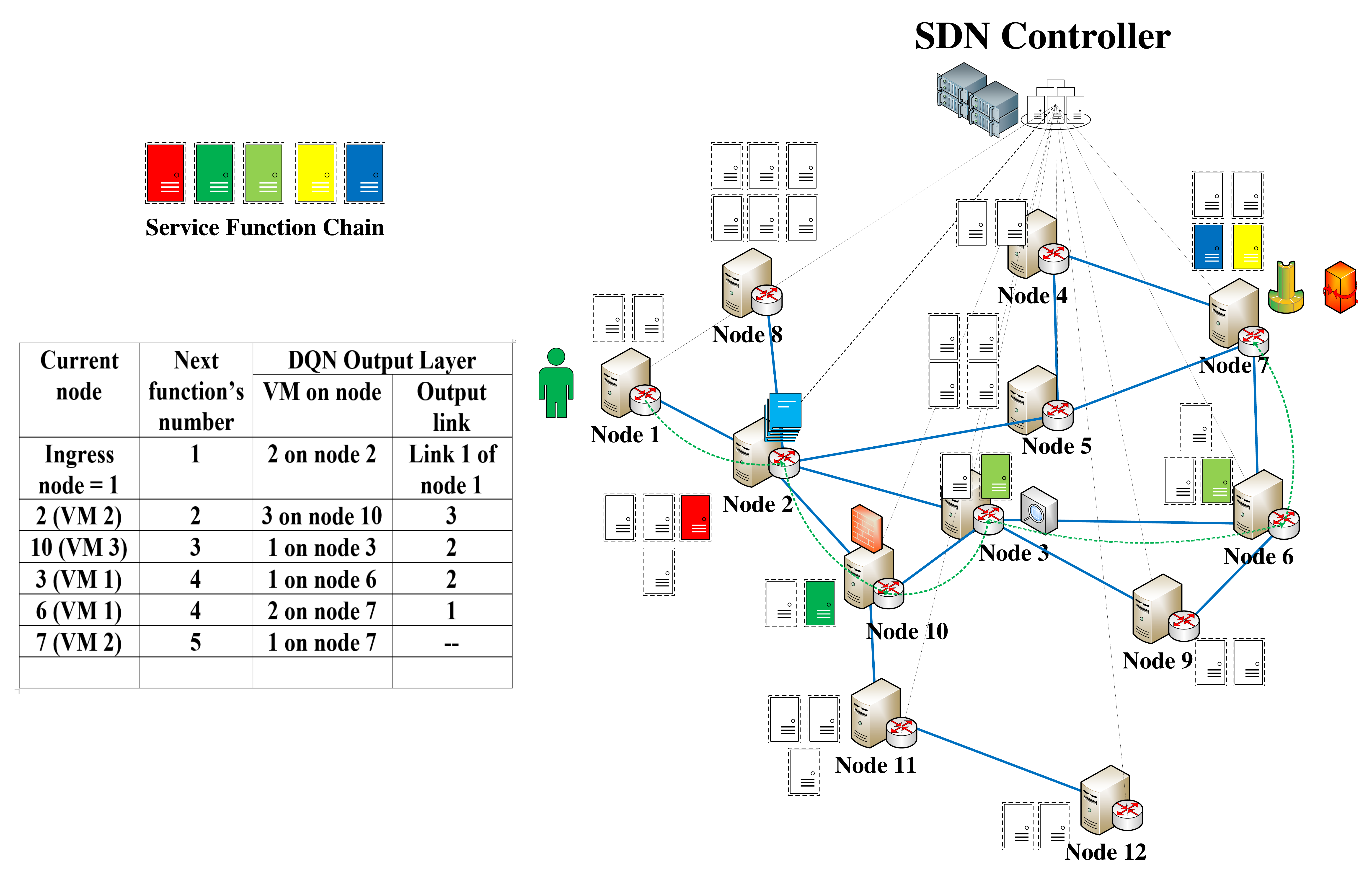}
\caption{An example of  function placement and node by node routing algorithm for a specific service}
\label{fig:sfc-and-routing}
\end{figure}

\begin{figure}
	\centering
	\includegraphics[width=0.5\linewidth]{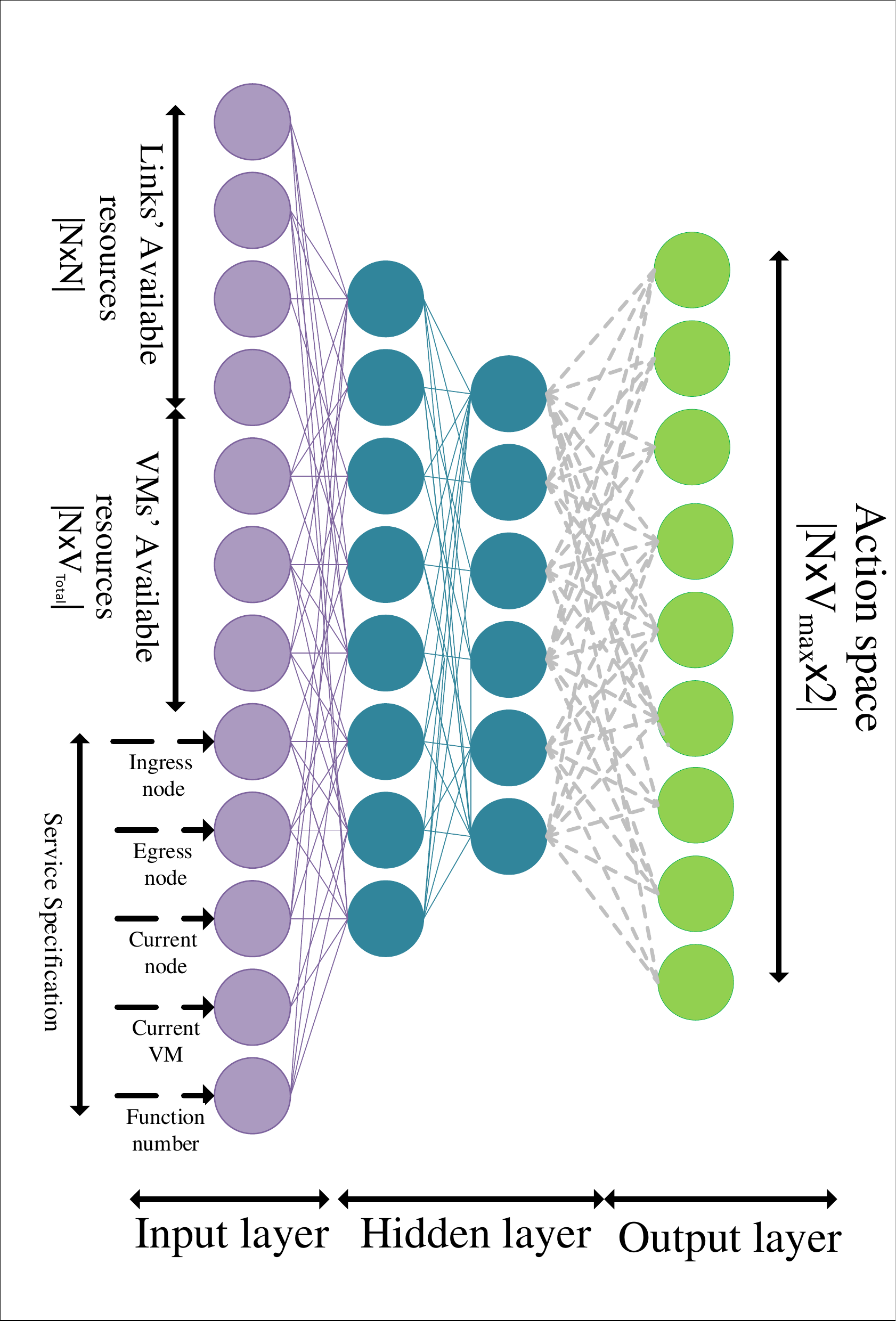}
	\caption{The DQN  based on the network state and service specification determines the action.}
	\label{fig:dnnout}
\end{figure}

\begin{algorithm}
	\tiny
	\DontPrintSemicolon
	\renewcommand{\arraystretch}{0.9}
	\caption{DQN Algorithm}
	\label{DQN}
	Initialize the DNN with random weights and consider the network graph and capacity of the links and VMs;  $\mathcal{G}, \boldsymbol{\Psi},   \bold{X}, \text{and set the initial weighth for DNN}:~\theta_{0}$ \\
	\If{the central controller want taking an action}{Action $a^{t,j}_{k}$ is selected;\\
		$  a^{t,j}_{k}= 
		\begin{cases}
		\text{ select a random action} &\text{with probability}~\epsilon,\\
		\argmax_{a}(Q(s^{t,j},a))&\text{with probability}~1-\epsilon
		\end{cases}$
		\\
		Store transaction ($s^{t,j},a^{t,j}_{k},r^{t,j}_{k},s^{t+1,j}$) in memory $\mathcal{\hat{D}}$\\
		Sample random minibatch with batch size $\hat{\mathcal{B}}$  of transitions $(s^{t,j},a^{t,j}_{k},r^{t,j}_{k},s^{t+1,j})$ from $\mathcal{\hat{D}}$}	
	\If{a request accepted}{Perform Algorithm \ref{Remain_Al}}
	\Else{Perform  gradient descent step on Mean Square Erorr (MSE)  of  $(r^{t,j}_{k}-Q(s^{t,j},a^{t,j}_{k};\theta ))^2$ and update parameter $\theta^{t}$ of network}	
\end{algorithm}
\textbf{Action Space:} The action space is denoted by $\mathcal{A}$ which includes all the network VMs on the nodes that can be considered for function placement or as a switch.
Based on the network state and SFC requirements, a subset of  actions is possible that is denoted by $\mathcal{A}_{p}\subset\mathcal{A}$. For example, if user $u$ requests service chain $\left\lbrace f^1 \rightarrow f^2 \rightarrow f^3 \right\rbrace$, the corresponding action determines that the next node and VM is selected for function placement or just it is a switch.
In fact, we propose a smart and  adaptive NFV-RA algorithm that  perform  joint function placement and node by node dynamic routing. More details are given in Algorithm \ref{3}. Thus, the size of  all action space for each of service request is calculated by
\begin{align}
|\mathcal{A}| = |\mathcal{N}|\times|\mathcal{V}_\text{Total}|\times 2.
\end{align}
Consequently, the agent for user $u$ in the service $k$ selects action $a^{t,j}_{k}\in \Bbb{N}$ at time slot $t$ in $j$th step of Algorithm \ref{3}.\\
$\bullet$ \textcolor{black}{\textbf{DQN-AR  for RA, Dynamic Routing,  and Function Placement:}} To adopt Algorithm \ref{DQN} for dynamic routing and function placement, we propose an algorithm that  by  an interactive approach with Algorithm \ref{DQN}, performs a node by node routing and function placement  beginning from $n_{i,k}$ and in each step of routing algorithm, considers the current node as $n_{c,k}$ and continues to reach  $n_{e,k}$. On the other hand, for each of service requests, with considering the network state and service specification as inputs of DQN, the output of the DQN determines the corresponding action as the next node and VM in SFC path. 
It is worth mention that in each step, only a set of the actions is possible. 
 We consider the set of  nodes that are directly connected to current node $n_{c,k}$  and it is denoted by $\mathcal{N}_{c,k}$. 
	 Subsequently, we consider the set of VMs in which they are on the set $\mathcal{N}_{n,c}$ as set of the possible actions and it is denoted by $\hat{\mathcal{N}}_{c,k}$.
	In addition, we assume that the agent can choose a VM form $\hat{\mathcal{N}}_{c,k}$.
	The agent can placed a function on the selected VM or consider the  selected VM as a forwarding device. Furthermore, we assume $\mathcal{A}_{p,p}=\mathcal{A}_{p,s}=\left\lbrace  \hat{\mathcal{N}}_{n_{c,k}} \right\rbrace $ where $\mathcal{A}_{p,p}$ and $\mathcal{A}_{p,s}$ are the sets of possible actions for function placement and  router selection, respectively.
Thus, the set of possible action is defined by $\mathcal{A}_{p}=\mathcal{A}_{p,p}\cup\mathcal{A}_{p,s}$.
To determine the type of each action, we define an auxiliary binary variable as $a^{t,k}_{s,k}$ as follows:
\begin{align}
a^{t,k}_{s,k}=\begin{cases}
0,~~ \text{If} ~~ a^{t,k}_{k}\equiv_2 0,\\
1,~~ \text{otherwise}.
\end{cases}
\end{align}
It is worth to mention that, if the agent chooses a possible action, $(a^{t,j}_{k}\in{\mathcal{A}}_{p})$, the
sub action $a^{t,j}_{s,k}$ determines the type of each action. Based on this, type of each action is defined by 
\begin{align}
	a^{t,j}_{k}\in
	\begin{cases}
		\mathcal{A}_{p,p}, & a^{t,j}_{s,k}=1,\\
		\mathcal{A}_{p,s}, & \text{otherwise}.
	\end{cases}
\end{align}
 In each step, if the  selected action is possible, then we check that this action belongs to which set.   If constraints \eqref{capa_cpu} and \eqref{capa_link} are satisfied, the function is placed on selected VM on corresponding node otherwise the request is rejected.  Similarly, for the links, we check the constraint \eqref{capa_link} sanctification. Nevertheless, the processing and propagation delay that incur the action $a^{t,j}_{k,u}$  is denoted by $\hat{\tau}$ and  calculated by
\begin{align}
	\hat{\tau} = 
	\begin{cases}
		\tau^{f,k}_{v,m} + i^{p_{m,m'}}_{n,n'}\rho^{k,e^{v,v'}_{m,m'}}_{p_{m,m'}}\kappa_{{n,n'}},     ~~~ &a^{t,k}_{s,j} =1,\\
		i^{p_{m,m'}}_{n,n'}\rho^{k,e^{v,v'}_{m,m'}}_{p_{m,m'}}\kappa_{{n,n'}},~~~ &\text{otherwise}.
	\end{cases}
\end{align}
In each step,  by checking constraints \eqref{delay}, we ensure the tolerable time of service request. 
\\$\bullet$~\textbf{Reward Function:} The agent after doing  action $a^{t,j}_{k}$ obtains a reward that is denoted by $r^{t,j}_{k}$ in $j$th step of the RA algorithm to service $k$ in time slot $t$. Nevertheless,  the agent selects a VM on a node for function placement or as forwarding device in $j$th step of Algorithm \ref{3}. Subsequently, if the  link between the current node, and the next node and  processing capacity of the next node's VM satisfy constraints \eqref{capa_link}, \eqref{capa_cpu} and \eqref{delay}, the agent obtains reward that is calculate by the following: 
\begin{align}
r^{t,j}_{k}=w_{acc}-w_{cost}\tilde{\phi}^{j}_{k}  ~~~ &, \forall k \in \mathcal{K}, \forall t,
\end{align} 
where $w_{acc}$ and $w_{cost}$ are coefficient factors of constraint satisfaction and cost and $\tilde{\phi}^{j}_{k}$ is the cost of the action, given below:
\begin{align}
&\tilde{\phi}^{j}_{k}= a^{t,k}_{s,k}w_{m,v}d^{k}_{f}B_{k}\delta^{k}_{u}\xi^{f,k}_{v,m}+
\\ \nonumber
&\hat{w}_{{n,n'}}B_{k}i^{p_{m,m'}}_{n,n'}\rho^{k,e^{v,v'}_{m,m'}}_{p_{m,m'}}, \forall u \in \mathcal{U}, \forall f \in \mathcal{F}_{k},\forall k \in \mathcal{K}.
\end{align} 
Otherwise, if the constraints are not satisfied, the request is rejected and the agent reward is set to $0$. Based on this the reward of each step $j$ is calculated by following:
\begin{align}
r^{t,j}_{k}=\begin{cases}
w_{acc}-w_{cost}\tilde{\phi}^{j}_{k},~~~ &\text{if the constraints  satisfied},
\\
0,  ~~~~~~~~~~ &\text{otherwise}.
\end{cases}
\label{reward_f}
\end{align}
In fact,  for each step $j$ of the Algorithm \ref{3}, we define a reward that depends on constraints sanctification and cost of each action. Finally, the total reward that the agent obtained  is defined by
\begin{align}
r^{t}_{k}=\sum_{j}r^{t,j}_{k}.
\end{align}
\begin{algorithm}
	\tiny
	\renewcommand{\arraystretch}{10}
	\caption{DQN-based NFV-RA algorithm for dynamic routing and function placement}
	\label{3}
	\For{each time slot $t$}{
	\For{each service $k$}{
	 $t_{u}\leftarrow$Arivial time, 
	$t_{o}\leftarrow0, \ i_{f}\leftarrow0$, $j\leftarrow0$, $\hat{p}=\left\lbrace \right\rbrace$,
	$n_{c,k}\leftarrow n_{i,k}$\\
	\While{Constraint \eqref{delay} is satisfied, $t_{o}<\tau_{k}$}{\While{$n_{c,k}\neq n_{e,k}$}
	{\While{$j<J$}{ \If{$a^{t,j}_{k,u}\in\mathcal{A}_{p}$}{ $\hat{p}\leftarrow \hat{p}\bigcup\left\lbrace a^{t,j}_{k}\right\rbrace $\\ \If{$a^{t,j}_{k}\in \mathcal{A}_{p,p}$}
	{\If{$i_f<F_{k}$}
		{\If{Constraints \eqref{capa_cpu} and \eqref{capa_link} are  satisfied}{Calculate $r^{t,j}_{k}$\\$i_f\leftarrow i_f+1$, $f\leftarrow f',j\leftarrow j+1, t_{o}\leftarrow t_{o}+\hat{\tau}$}\Else{request is rejected\\$r^{t}_{k}\leftarrow0$}}
    \Else{consider $a^{t,j}_{k}$ as switch\\
    $j\leftarrow j+1,t_{o}\leftarrow t_{o}+\hat{\tau}$}
}
\Else{\If{Constraint \eqref{capa_link} is satisfied}{Calculate $r^{t,j}_{k}$\\$j\leftarrow j+1,t_{o}\leftarrow t_{o}+\hat{\tau}$}}}}
\Else{request is rejected\\$r^{t}_{k}\leftarrow0$
}
}
\If{$i_f<F_{k}$}{\If{Constraint \eqref{capa_cpu} and $i_f<F_{k}$}{Function $f$ is placed on engress node;\\Calculate $r^{t,j}_{k}$\\$i_{f}\leftarrow i_{f}+1$, $f\leftarrow f'$, $j\leftarrow j+1,t_{o}\leftarrow t_{o}+\hat{\tau}$}}}\If{$t_o>\tau_{k}$}{request is rejected\\$r^{t}_{k}\leftarrow0$
}
}
}
\end{algorithm}
The designed DQN is depicted in Fig. \ref{fig:dnnout}. According to the figure, by considering the network state as DQN input, the DNN output layer determines the actions.
 Note that some of the nodes in the path are only forwarding devices (e.g., switch) (see Fig. \ref{fig:SFC} and \ref{fig:sfc-and-routing}).
\section{COMPUTATIONAL COMPLEXITY}\label{section_complexity}
\textcolor{black}{
We analyze the computational complexity of the proposed DQN-AR algorithm and then we compare it with the NFV deep algorithm \cite{10.1145/3326285.3329056}, Tabu search algorithm \cite{leivadeas2019vnf}, and greedy search algorithm which is the well know algorithm that is deployed in \cite{sheikhzadeh2021ai, li2020cost ,agarwal2018joint}.} The complexity of DNN based algorithms is depended on the architecture, configuration, number  of input and output, and hidden layers. Moreover, for deploying DNN in the DQN-AR algorithm,  considering the action space size and state space size is required \cite{10.1145/3326285.3329056}. Also, by considering the number of output layer neurons as $P_o$, number of the input layer neurons as $P_i$, and number of the hidden layers as $P_h$, the time complexity of the proposed DQN-AR for each action is obtained by following:
\begin{align}
\mathcal{O}(P_n\times (P_i+ P_h \times P_n + P_o)),
\end{align} 
where $P_n$ is the hidden layer's neuron number \cite{pei2019optimal}. Also, as can be seen from Fig. \ref{fig:dnnout}, $P_{i}=|\mathcal{L}|+|\mathcal{N}|\times|\mathcal{V}_{\text{Total}}|+5$ and $P_o=|\mathcal{N}|\times|\mathcal{V}_{\text{Total}}|\times2$.
Moreover, by considering $E$ iterations in the case of Tabu search, for $|\mathcal{K}|$ number of service requests, the time complexity is obtained by $\mathcal{O}(E\times |\mathcal{K}|\times F)$ for $F$ number of functions in a certain SCF.
Accordingly, by increasing the number of iterations, the complexity of the Tabu search is increasing that can cause more complexity in the case of problems with a larger space of feasible solutions.
Finally, to find the shortest path from ingress node to egress nodes for each of services with $F$ functions, in a network with $|\mathcal{N}|$ nodes  and $|\mathcal{L}|$ links, the total time complexity is  obtained by $\mathcal{O}(|\mathcal{K}|\times(|\mathcal{L}| + |\mathcal{N}|\log(|\mathcal{N}|)+ F\times K)$.
\section{SIMULATION RESULTS}\label{Simulation_Results}
\textcolor{black}{We analyze the  performance of the proposed method  using simulations. Accordingly, first we investigate the convergence of the proposed method. Next, we evaluate the effect of the coefficient factors in the the objective function. Afterwards, we compare the  results of the proposed method with the baselines.}
\subsection{{Simulation Setup}}
\begin{table}[]
	\caption{Simulation Setting}
	\centering
	\label{tab:setting}
	\scalebox{.6}{
	\begin{tabular}{|c|c|}
		\hline
		\rowcolor[HTML]{00D2CB} 
		Parameters                               & Value                                                                                                                                               \\ \hline
		& \cellcolor[HTML]{96FFFB}Average Duration Time:                                                                                                      \\ \cline{2-2} 
		& 240, 600, 900, 1200 seconds  \cite{pei2019optimal}                                                                                                                       \\ \cline{2-2} 
		& \cellcolor[HTML]{96FFFB}Data Rate:                                                                                                                  \\ \cline{2-2} 
		& Max = 4 Mbps   Min = 64Kbps      \cite{savi2019impact}                                    \\ \cline{2-2} 
		& \cellcolor[HTML]{96FFFB}Average Tolerable Time                                           \\ \cline{2-2} 
		& Max = 500ms    Min = 100ms                                                                                                                               \\ \cline{2-2} 
		& \cellcolor[HTML]{96FFFB}VNF and Services:                                                                                                                        \\ \cline{2-2} 
		\multirow{-8}{*}{Service Specification}  & FW, NAT, IDNS, TM, VOC      \cite{savi2019impact}    \\ &  Web Browsing, Voice over IP, Video Streaming                                                 \\ \hline
		& \cellcolor[HTML]{96FFFB}VM's Capacity:                                                                                                              \\ \cline{2-2} 
		& \begin{tabular}[c]{@{}c@{}}Max = 1200 CPU Cycle per second\\ Min = 200 CPU Cycle per second\\ \cite{ebrahimi2020joint}\end{tabular} \\ \cline{2-2} 
		& \cellcolor[HTML]{96FFFB}Link's Capacity:                                                                                                            \\ \cline{2-2} 
		\multirow{-4}{*}{Network Resources}       & \begin{tabular}[c]{@{}c@{}}Max = 6400 Mbps\\ Min = 1600 Mbps\\ \cite{ebrahimi2020joint}\end{tabular}                               \\ \hline
		& \cellcolor[HTML]{96FFFB}Number of the Server Node:                                                                                                  \\ \cline{2-2} 
		& 10, 20, 30, 50, 100     \cite{10.1145/3326285.3329056}                                                    \\ \cline{2-2} 
		& \cellcolor[HTML]{96FFFB}Propagation delay on the links                                                                                              \\ \cline{2-2} 
		& Max = 15ms   Min = 5ms                                                                                                                                  \\ \cline{2-2} 
		& \cellcolor[HTML]{96FFFB}Number VMs of each nodes                                                                                                    \\ \cline{2-2} 
		\multirow{-6}{*}{Network  Configuration} & $V_\text{max}$=6                                                                                                                                                   \\ \hline
	\end{tabular}
}
\end{table}
As listed in Table. \ref{tab:setting}, we  consider some of the service specifications based on  their QoSs \cite{savi2019impact} and service lifetime.
We assume that each time slot is equal to one second. We consider $1000$ to $6000$ time slots for  the simulation time, and $2000$ iterations \cite{10.1145/3326285.3329056} with $10$ Monte Carlo repetitions. Moreover, we generate the number of service requests by the Uniform random process \cite{ding2019deep} and the service life time by the exponential random process. Also, to set the ingress and egress nodes for set $\mathcal{K}$, at the beginning of the simulation, we select some random nodes among the network nodes. 
\\\indent
Subsequently, to have a network with certain number of  edges and nodes, we generate a random connected graph through \textit{NetworkX} libraries in Python \cite{fu2019dynamic}, \cite{ebrahimi2020joint}. Also, to deploy DNN, we use \textit{Tensorflow} and \textit{Keras} libraries in Python.  Moreover, for the cost weight, we consider  $w_{n,v}$ and $\hat{w}_{n,n'}$ in range of $25$ to $75$ \$/Mbps \cite{gholipoor2020e2e}. In addition, the source code of the proposed DQN-AR is available in \cite{ali2021}.
\subsection{{Simulation Results Discussions}}
\textcolor{black}{We evaluate the effect of the main parameters, such as, services' life time,  number of the server nodes of the network (network topology), and the number of the arrival service requests on different  baseline algorithms.}
\subsubsection{Average Acceptance Ratio (AAR)} 
\begin{figure}
	\centering
	\includegraphics[width=1\linewidth]{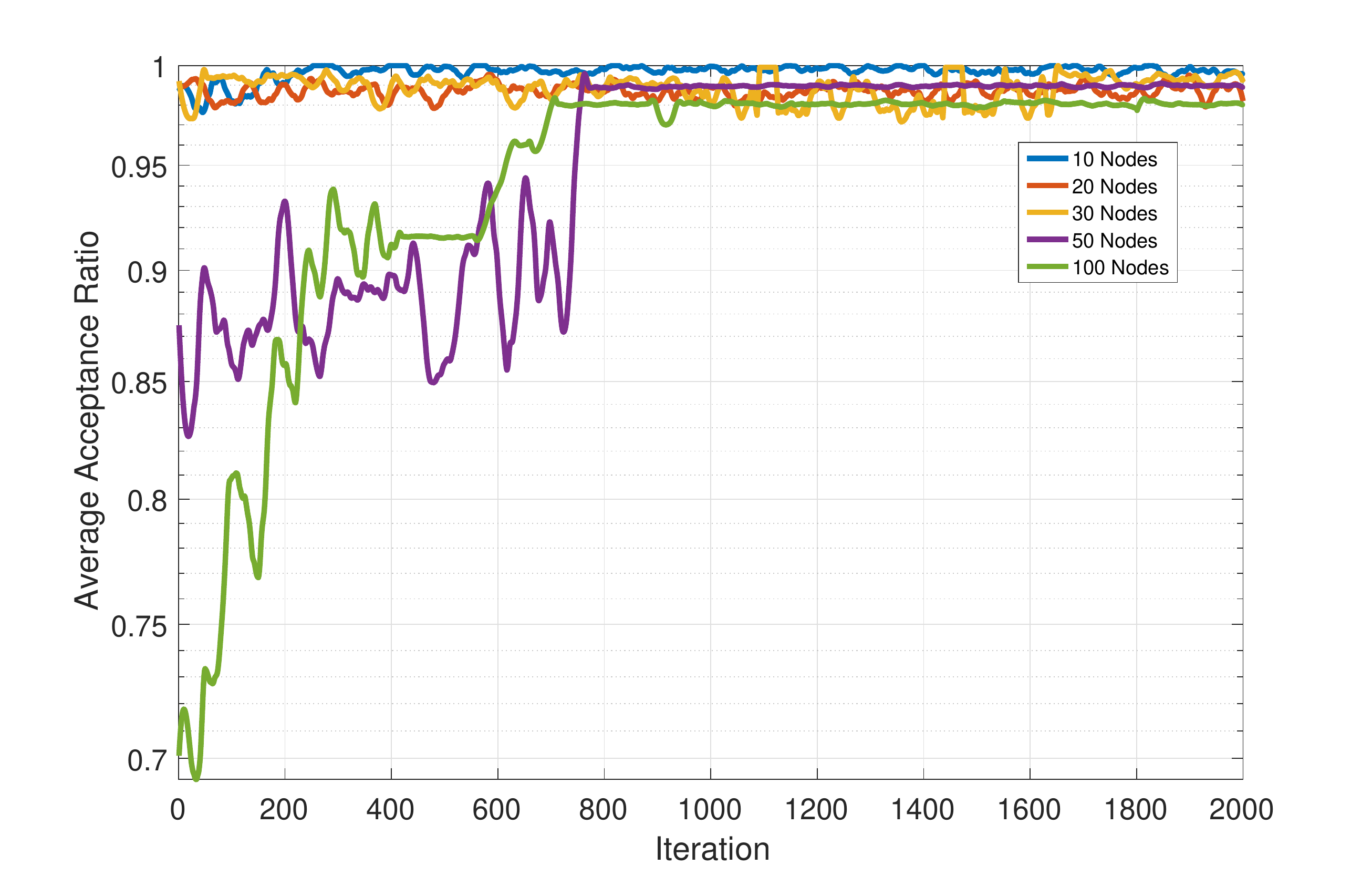}
	\caption{AAR over the iterations for different network topology}
	\label{fig:10-100acc}
\end{figure}

As the network topology, such as the number of the nodes and links and their configurations, has a significant effect on the routing algorithm and protocols, we evaluate the AAR on different network typologies. To have a comparison of  the effect of the network topology on the performance of the agent, we consider the networks with size 10 to 100 nodes to evaluate the AAR over the iteration number.
As can be seen in Fig. \ref{fig:10-100acc}, in the first iteration, the  AAR for the different network typologies have significant differences, specially, in the networks with large number of the server nodes. It is  because that in a large network, the agent  needs to select more actions to find  appropriate path from ingress node to egress node and also the SFC placement on the VMs for each service request. Gradually, the AAR  increases over the iterations. That is because the agent learns how to handle the requests and find the appropriate path from ingress nodes into egress nodes in different states. 
\subsubsection{Average Network Utilization Cost (ANUC)} \label{ANUC1}
\begin{figure}
	\centering
	\includegraphics[width=1\linewidth]{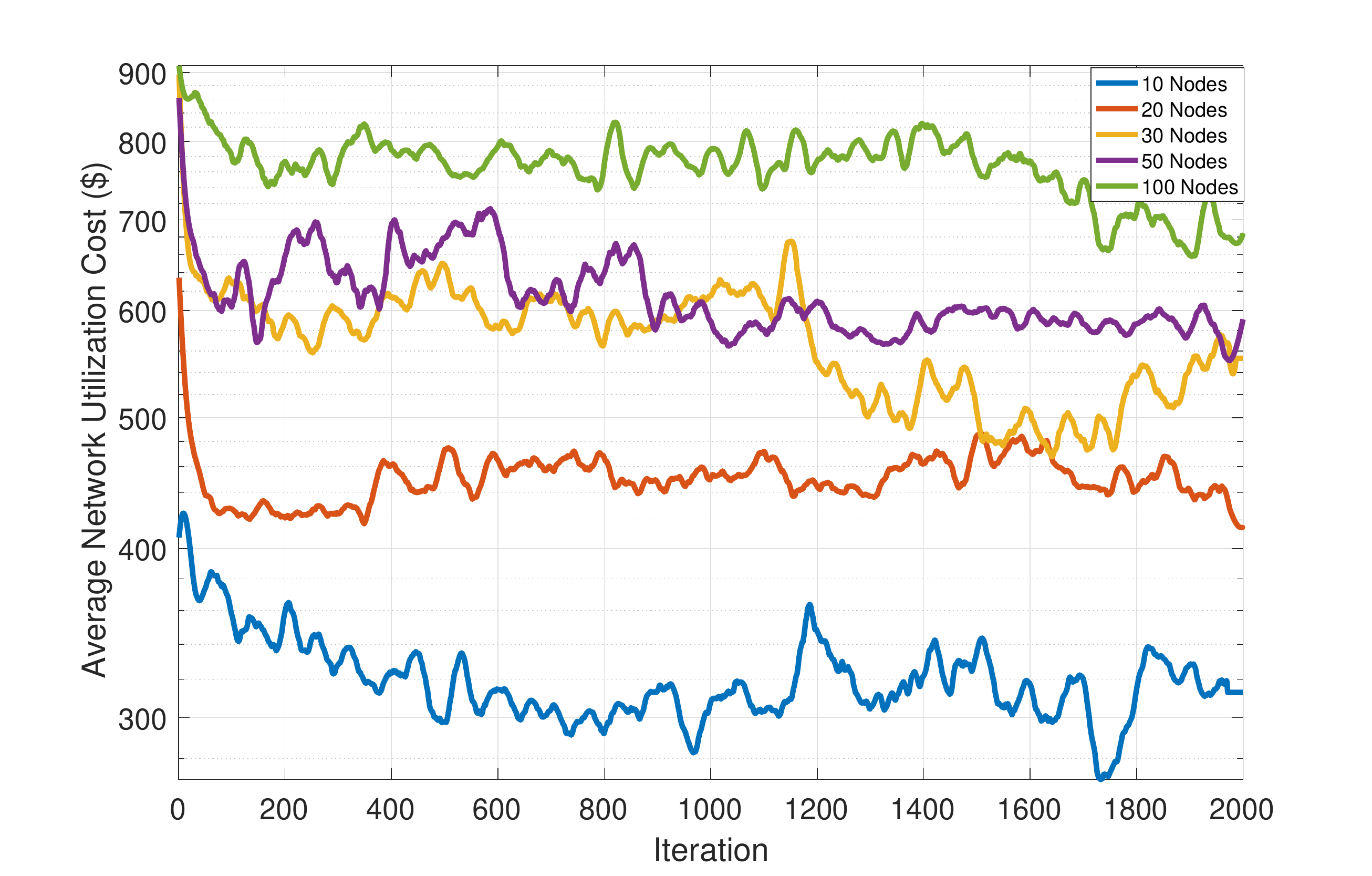}
	\caption{ANUC over the iterations for different network topologies}
	\label{fig:10-100cost}
\end{figure}

Network topology and configuration have a significant effect on the length of the paths. To evaluate the effect of the network topology on the ANUC, we consider the network with 10 to 100 nodes.
Because of the significant differences in the length of the paths from ingress to egress nodes in small and big networks, ANUC  depends on the network size as shown in Fig. \ref{fig:10-100cost}.
Since the initial actions are selected by the agent randomly, we see that the obtained utilization cost is very high. After that, the agent gets more experience and take the actions based on the obtained experience and  the ANUC
gradually decreases over the iterations.

\subsection{Baselines Algorithms} 
 In order to evaluate the performance of the proposed DQN-AR, we consider  baselines for comparing the results for different setting. Since DQN-AR is an online and adaptive algorithm in routing and function placement, it shows good performance in  different conditions. To evaluate the performance of the proposed algorithm, we  consider \textit{NFVdeep} as baseline $1$, Tabu search algorithm as baseline $2$, and greedy algorithm as baseline $3$, that are studied in \cite{10.1145/3326285.3329056}, \cite{leivadeas2019vnf}, and \cite{agarwal2018joint},  respectively.
\subsubsection{Effect of average number of the requests over time}
\begin{figure}
	\centering
	\includegraphics[width=1\linewidth]{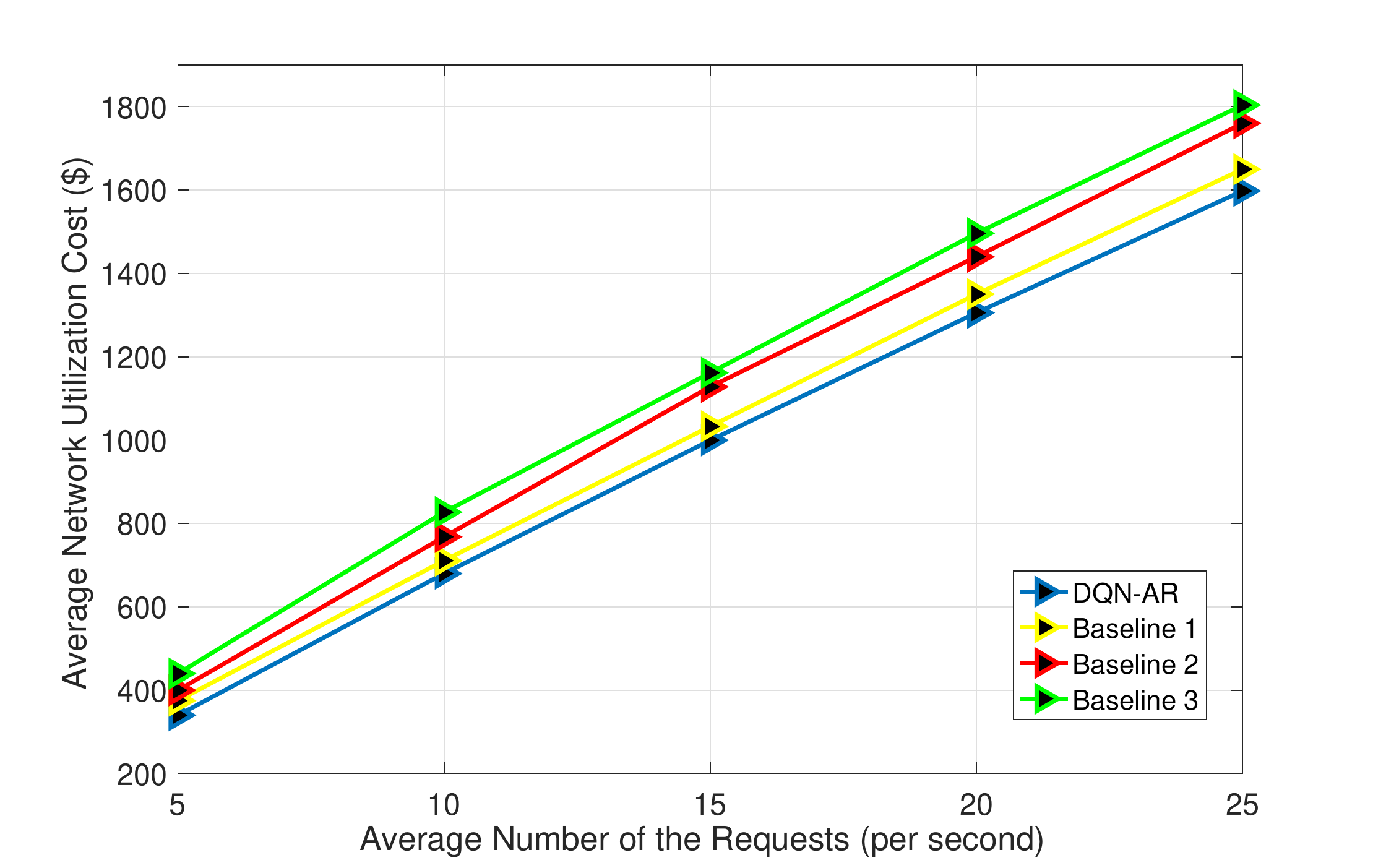}
	\caption{Average network utilization cost versus the average number of requests per second}
	\label{fig:cost_req}
\end{figure}

\textcolor{black}{To analyze the effect of the number of requested services on the ANUC, we  increase the average number of users from 5 to 25 requests per second. As can be seen in Fig. \ref{fig:cost_req}, by increasing the number of arrival services, ANUC increases. By deploying adaptive function placement  and dynamic routing in the proposed DQN-AR, we obtain  lower ANUC for different number of arrival services.}
\\

\subsubsection{Effect of the coefficient $w_{\text{cost}}$ on AAR}  
\begin{figure}
	\centering
	\includegraphics[width=1\linewidth]{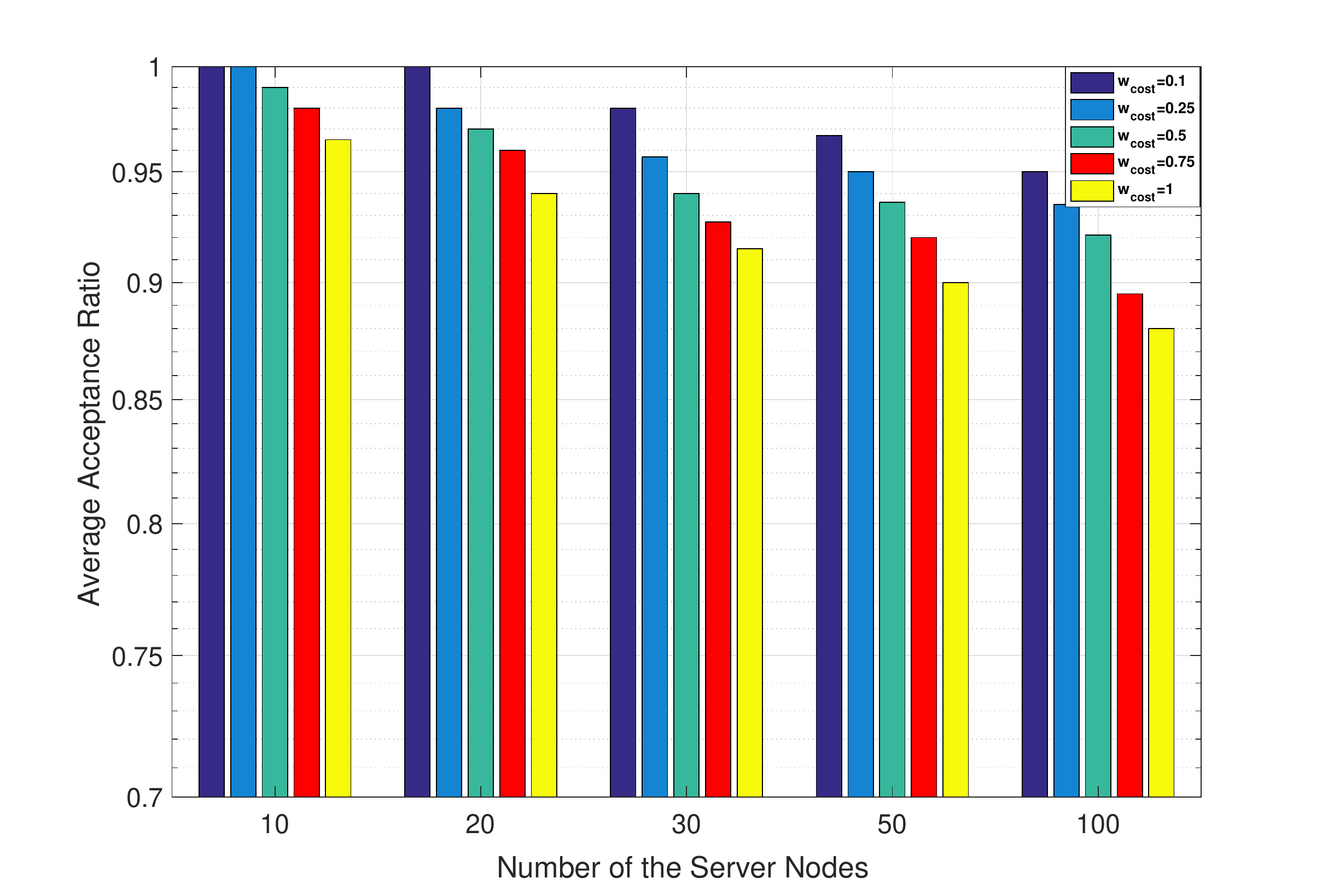}
	\caption{Comparing AAR with considering different coefficient of the cost in reward function}
	\label{fig:accalpha}
\end{figure}

The ANUC is very dependent on the AAR, since when the accepted requests  increases, the network utilization cost increases simultaneously. Based on this, we try to maximize the number of accepted requests  with respect to the constraints and minimize the utilization cost at the same time. Also, as we denote in \eqref{reward_f}, we consider the reward function  with certain coefficients as $w_{\text{acc}}$ and $w_{\text{cost}}$. Accordingly, the coefficient $w_{\text{cost}}$ determines the priority of cost in each action.
As we show in Fig. \ref{fig:accalpha}, by increasing the coefficient $w_{\text{cost}}$, the AAR  deceases.
\subsubsection{Effect of the coefficient $w_{\text{cost}}$ on ANUC}
\begin{figure*}
	\centering
	\includegraphics[width=1\linewidth]{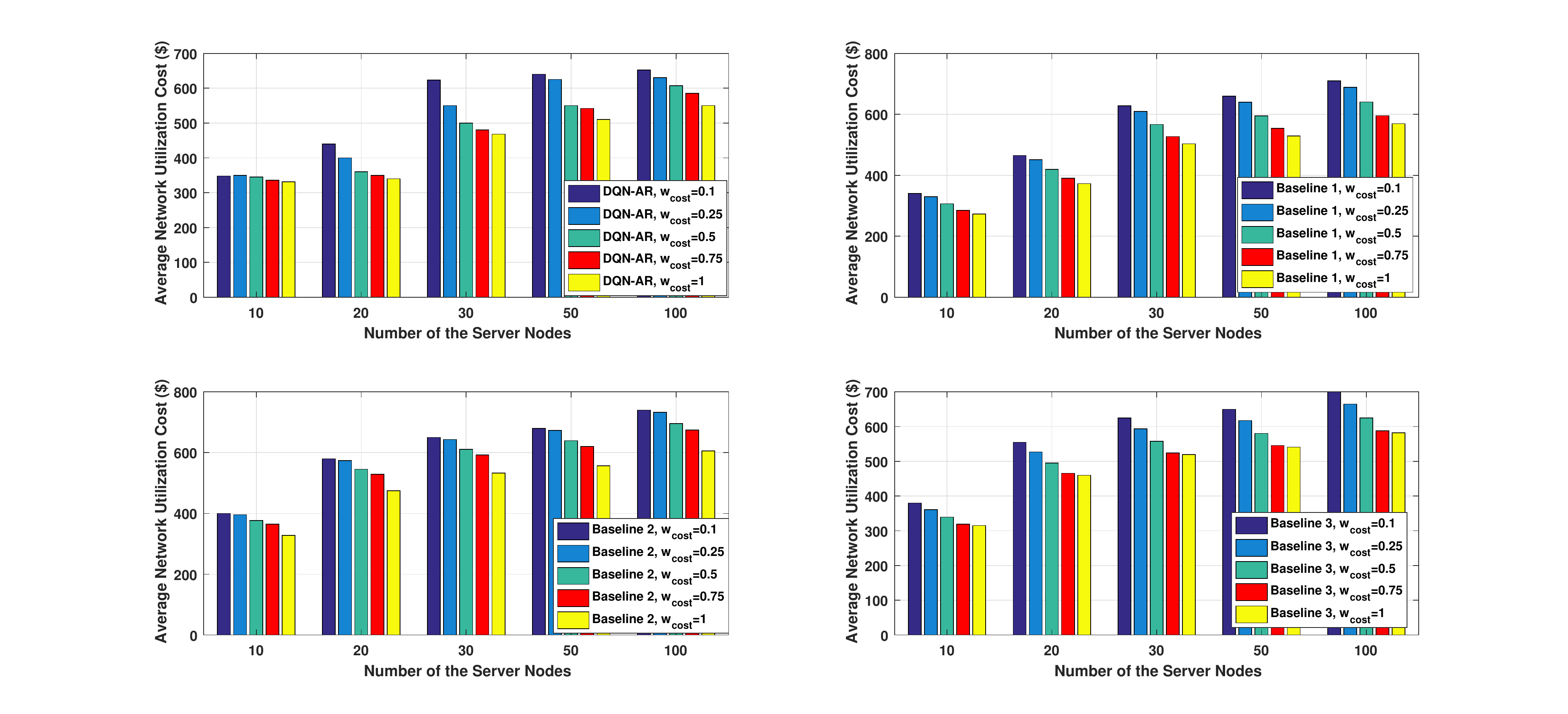}
	\caption{Performance evaluation of the proposed method and the baselines by changing the coefficient $w_{\text{cost}}$}
	\label{fig:w_cost}
\end{figure*}

By considering the coefficient $w_{\text{cost}}$, the agent has more attention to minimize the ANUC. Therefore, the agent chooses actions
that have less cost, but these actions can not provide sufficient resources for the next requests Fig. \ref{fig:w_cost}. Because ANUC is closely dependent on the AAR, by decreasing AAR, ANUC gradually decreases, but by considering this coefficient, AAR decreases 12\%  and ANUC decreases 20$\%$ in the proposed DQN-AR method.
\textcolor{black}{In addition, to evaluate the effect of coefficient $w_\text{cost}$ on the baselines, we illustrate the obtained results in Fig. \ref{fig:w_cost}. Baseline 1, by placing the VNF in the VMs by the \textit{NFVdeep} algorithm achieves more ANUC compared to the proposed method. Baseline 2 deploys Tabu-search algorithm for function placement and routing and achieves higher cost than baseline 1. Finally, baseline 3, by deploying greedy-based selection criteria, has the worst results specifically in the case of large networks.}
\subsubsection{Effect of Average Service Life Time on AAR}
\begin{figure}
	\centering
	\includegraphics[width=1\linewidth]{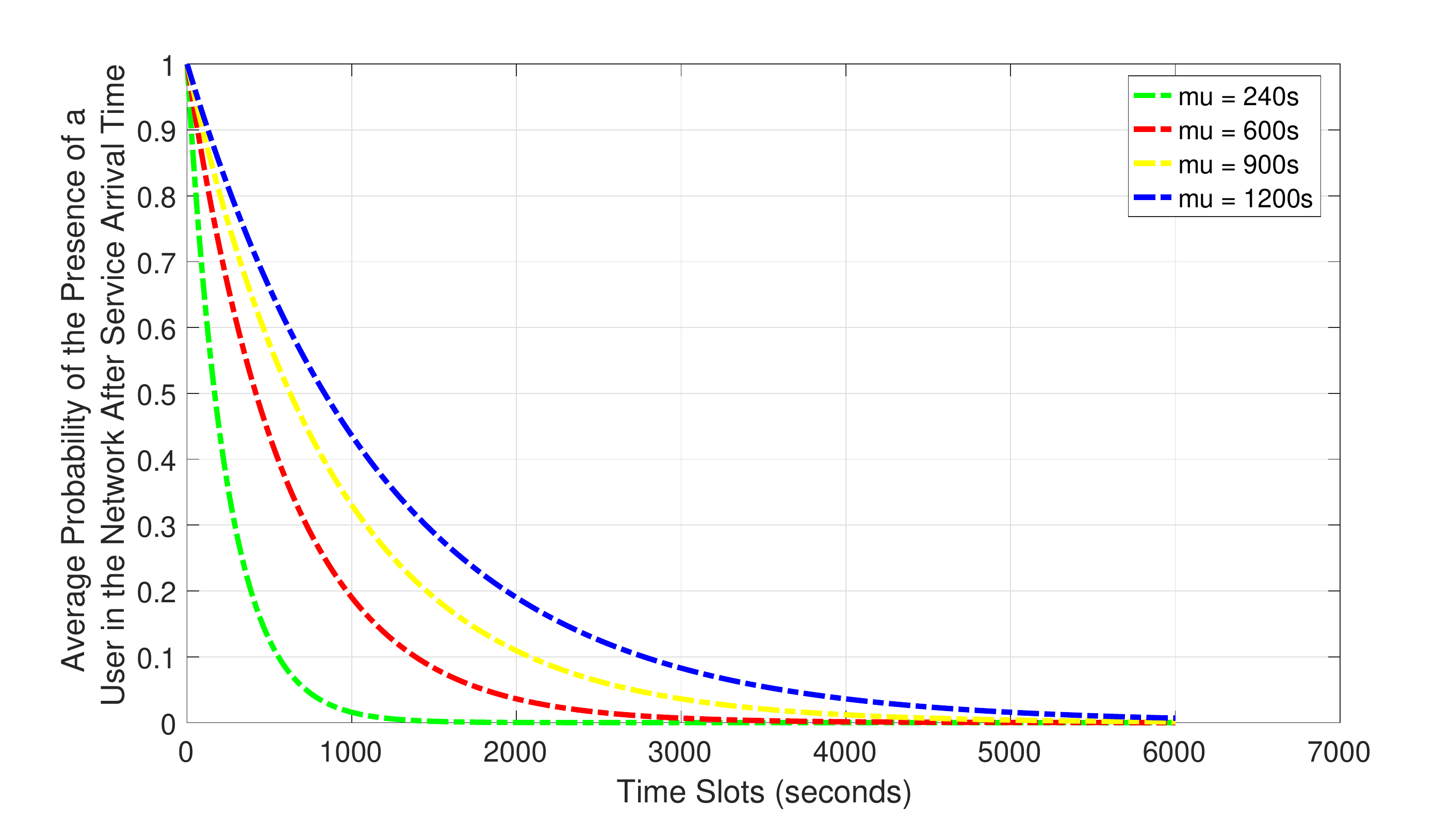}
	\caption{Average probability of the presence of a user in the network after service arrival time. }
	\label{fig:servicetime}
\end{figure}

\begin{figure}
	\centering
	\includegraphics[width=1\linewidth]{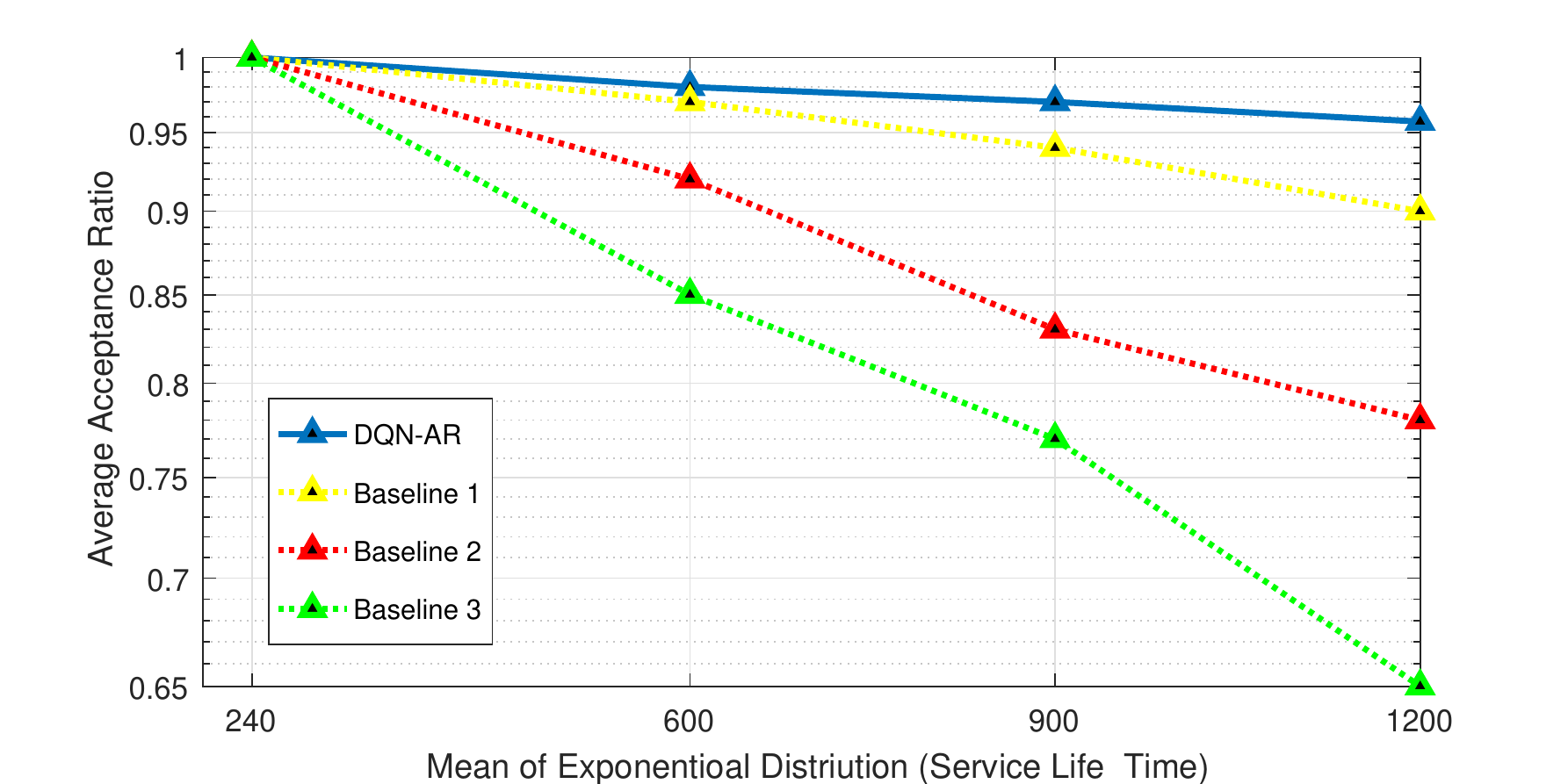}
	\caption{The effect of the service life time on AAR}
	\label{fig:mu}
\end{figure}

Average service life is a parameter that depends on the type of services.
To evaluate the effect of the service life time on AAR, we consider the service life time with 240 to 1200 seconds.
As can be seen in Fig. \ref{fig:mu}, increasing the services lifetime has more effect on AAR compared
to the number of requests. This is because when service lifetime becomes large, the available resource  decreases.  In addition, by considering the exponential distribution for the users' service lifetime, after a period of time equal to the mean of exponential distribution from the users' arrival time, as can be seen in Fig. \ref{fig:servicetime},  only 36$\%$ of these users departure the services.
\textcolor{black}{Because  effective resource  allocation according to the service specification have a significant effect on the AAR, DQN-AR by considering network resources and the service specification in the network state can adapt to the conditions that the available resources of network is limited. In addition, DQN-AR by performing an adaptive resource allocation, and dynamic routing achieves better results than baselines.}

\begin{figure}
	\centering
	\includegraphics[width=0.7\linewidth]{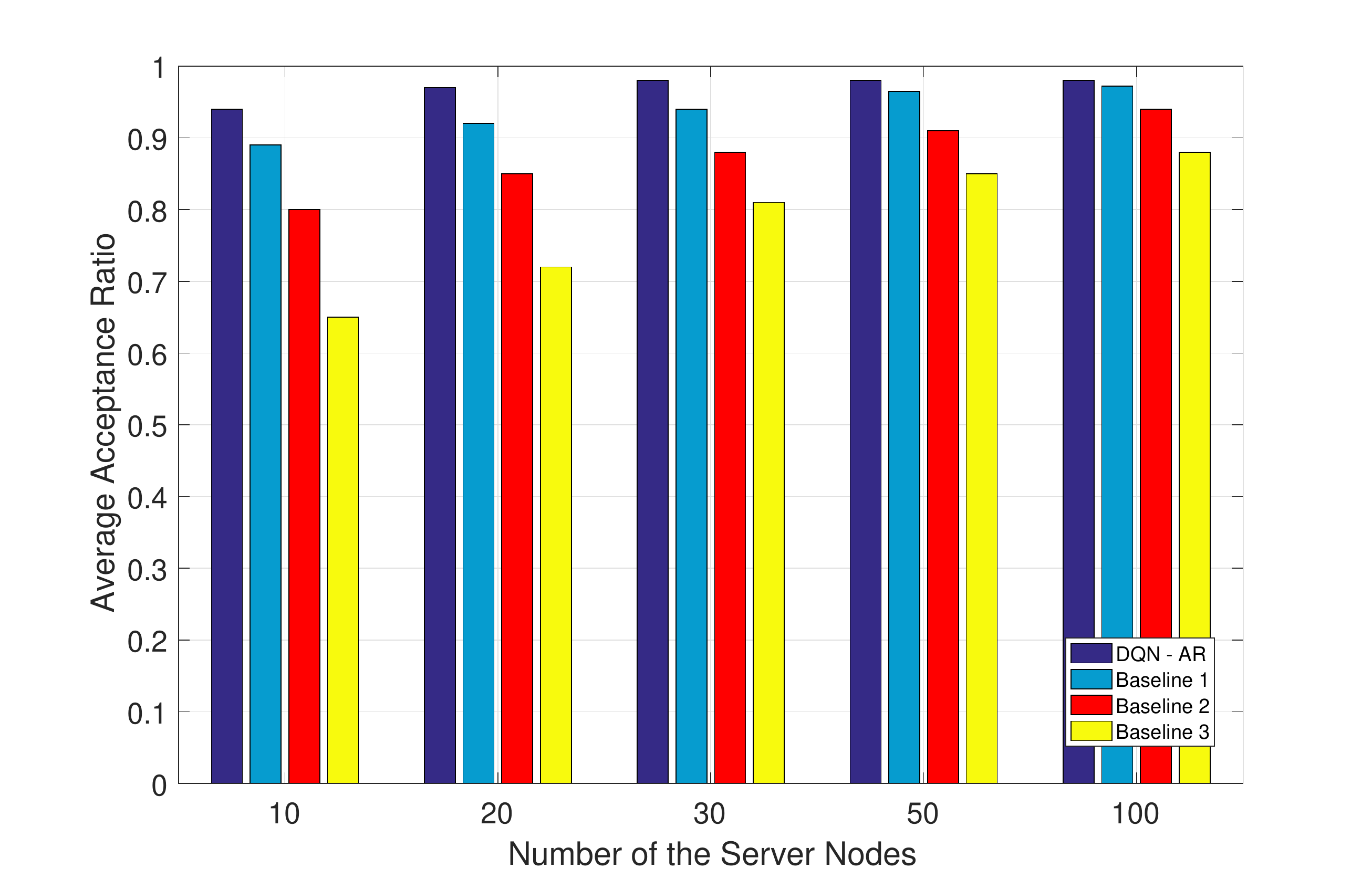}
	\caption{The effect of the network resources on AAR}
	\label{fig:acc_node}
\end{figure}

\subsubsection{Effect Network Resources on AAR}
To evaluate the effect of the available network resource on AAR, we consider that the users have maximum (1200 seconds) service life time. As can be seen in Fig. \ref{fig:acc_node}, by increasing the server nodes and links, the available resources  increases and the agent can accept more service requests. Because the proposed  DQN-AR algorithm can consider some of  the nodes as switch or for function placement and also deploy a dynamic node by node routing,  it has higher AAR in different network typologies.
\begin{figure}
	\centering
	\includegraphics[width=1\linewidth]{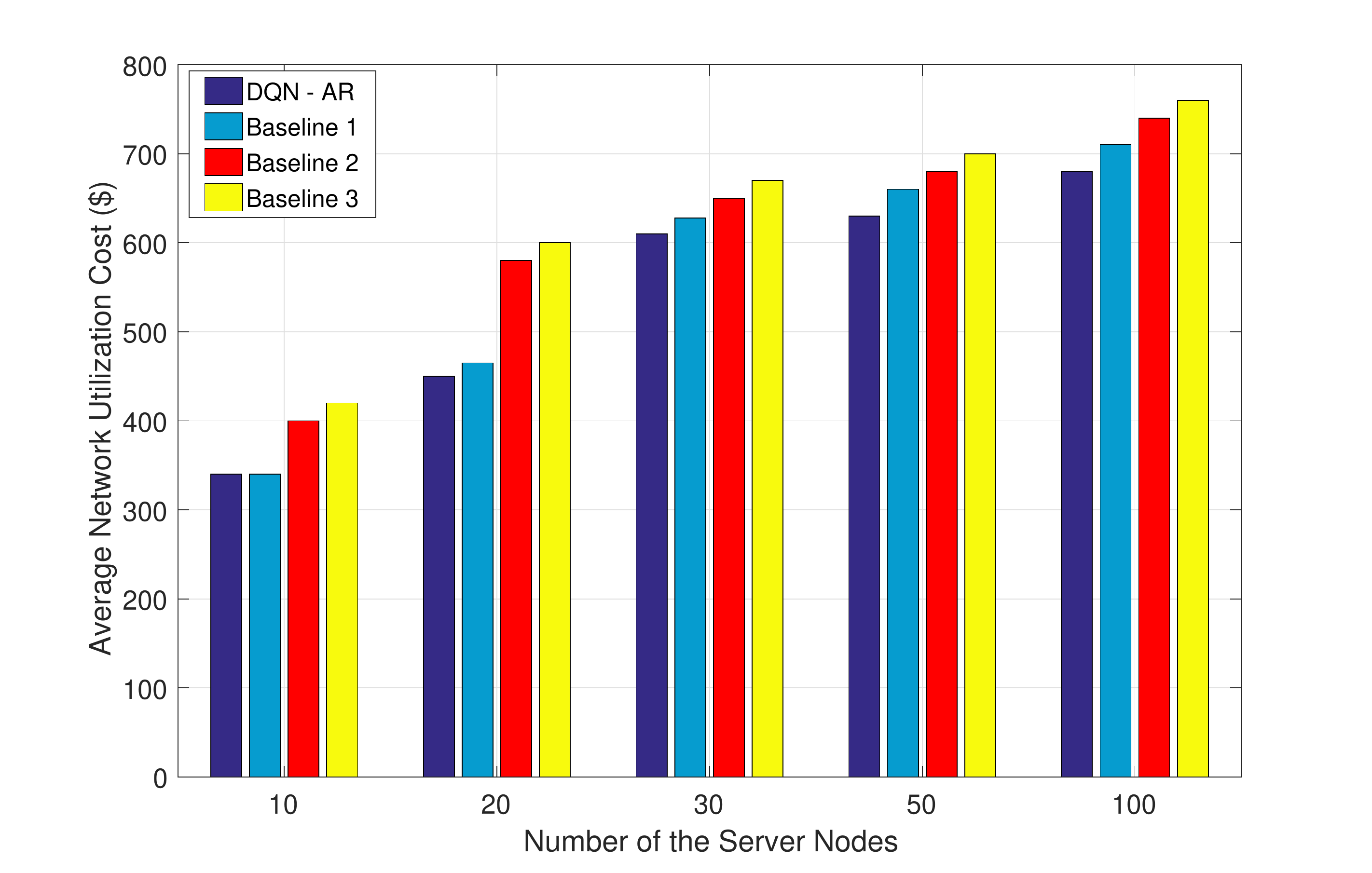}
	\caption{Effect of the network size and number of the server nodes on ANUC}
	\label{fig:cost_node}
\end{figure}
\subsubsection{Effect of the network topology on ANUC}
As we evaluated in Section \ref{ANUC1},
by increasing the network size and the number of the server nodes, because  the paths become longer, the AUNC is increased as shown in
Fig. \ref{fig:cost_node}. In fact, by increasing the server nodes, the network becomes bigger and also more scattered.
By solving the routing and function placement jointly in the proposed DQN-AR algorithm,  the ANUC is less than  that of the other baselines. In summery, since DQN-AR is an online
and adaptive algorithm in routing and function placement, it
shows good performance in different conditions.
\section{Future works}\label{Future_work}
\textcolor{black}{It will be important that future researches investigate the performance of the new RL-based methods that deploy combined methods like Recurrent Deterministic Policy Gradient (RDPG) to provide proactive and predictive resource allocation algorithms in NFV-enabled networks.  Therefore, in future works, we will study other RL-algorithms in NFV-enabled networks.}
\section{Conclusion}\label{Conclusion}
We studied an online service  provision framework by considering lifetime for each service and using  RA approach in a NFV-enabled network.
To this end, we formulated the cost of the network resource utilization for function placement and routing of the requested services by considering services requirements and resource constraints. To minimize the resource utilization cost by maximizing the service acceptance ratio, we defined the reward as a piecewise function.
Because of the large number of actions and states space, we used a DQN structure. 
Simulation results show the effectiveness of the proposed model. By evaluating the baselines, the network utilization cost is decreases by $5$ and $20\%$ and average number of admitted request increases  by $7$ up to $20$\%.
\bibliographystyle{ieeetr}
\bibliography{citationMECLearning}
\end{document}